\documentclass[10pt,journal,letterpaper]{IEEEtran}

\usepackage{amsmath,amssymb,amsfonts}
\usepackage{array}
\usepackage{booktabs}
\usepackage{cite}
\usepackage{graphicx}
\usepackage{multirow}
\usepackage{placeins}
\usepackage{url}

\begin{document}

\bstctlcite{BSTcontrol}

\title{Content-Edited Layout Generation for Graphic Design}

\author{Shirong Yang, Zheyi Pan, and Ying Cao%
\thanks{Shirong Yang, Zheyi Pan, and Ying Cao are with the School of Information Science and Technology, ShanghaiTech University, 393 Middle Huaxia Road, Pudong New Area, Shanghai, China (e-mail: yangshr2024@shanghaitech.edu.cn; panzhy2024@shanghaitech.edu.cn; caoying59@gmail.com).}%
\thanks{Corresponding author: Ying Cao.}}

\maketitle

\begin{abstract}
   In this paper, we formulate a novel form of conditional layout generation
   for graphic design, termed \emph{content-edited layout generation} (CELG).
   CELG extends conventional content-aware layout generation (CLG), which
   generates a foreground layout for a fixed background image. The key
   distinction is that CELG treats the input image as an editable visual asset
   and, conditioned on a textual design intention and a foreground requirement
   that may be empty, produces an adapted background together with its layout. To study
   this task, we introduce \emph{Search-Compare-Edit} (SCE), a data construction
   framework that supplies the paired supervision missing from existing design
   datasets. Starting from professional design canvases and their layouts, SCE
   synthesizes plausible pre-edit source images and corresponding editing
   instructions, yielding \emph{InstructEdit}, a dataset of aligned source
   images, target canvases, instructions, and layouts. Based on the generated
   data, we train a two-flow multimodal flow matching model that couples a
   pretrained image editor with a layout flow transformer, enabling background
   editing and layout generation simultaneously within a unified framework that
   more closely reflects real-world design workflows. Our experiments
   systematically demonstrate the value of CELG as a distinct task and the
   performance advantages of the proposed model.
\end{abstract}


\section{Introduction}
\label{sec:intro}

\IEEEPARstart{L}{ayout} creation is a fundamental task in graphic design, which is concerned with generating spatial arrangement of visual elements such as images, texts, embellishments on a canvas. Layout generation for graphic design has been widely studied, with numerous generative models proposed for both unconditional and conditional scenarios~\cite{li2024layoutditexploringcontentgraphicbalance,Hsu-IJCAI2025-ScanandPrint,horita2024retrievalaugmented,inoue2023layoutdm,gupta2021layouttransformerlayoutgenerationcompletion}. Recently, the community has witnessed a growing interest in content-aware layout generation (CLG)~\cite{zheng2019content, cao2022geometry, Hsu-2023-posterlayout,liu2025unilaydiffunifieddiffusiontransformer,tian2025relayoutintegratingrelationreasoning,hsu2025posterostructuringlayouttrees}, which aims to generate plausible and diverse layouts of elements on top of a given background image, and can find many use case scenarios in practice, such as creating advertising posters of commercial products.

However, CLG relies on a restrictive assumption about its input: the background image is treated as a ready-to-use canvas whose composition, visual appearance, and empty space have already been optimized for foreground elements. In practice, users often begin with an unprocessed visual asset, such as a product photograph, stock image, or rendering, rather than a professionally prepared background. Such assets may position the main subject poorly, contain distracting content in potential text regions, or provide insufficient negative space for headlines and other foreground elements.

A straightforward solution is to edit the raw asset first and then generate a layout on the resulting canvas. This sequential strategy, however, overlooks the interdependence between the two stages. Image editing determines the available space, visual saliency, and composition that constrain element placement, while the intended layout indicates where the canvas should be cleared, extended, or reorganized. Because the background is edited before the layout is known, problems such as insufficient text space or overlap between the subject and foreground elements may become apparent only after the layout has been generated. Correcting these problems requires editing the background again and regenerating the layout. Such iterative refinement increases inference cost without providing a direct mechanism for image editing and layout generation to coordinate. Conventional CLG therefore does not fully address design from raw assets, where canvas preparation and layout generation need to be considered together.

\begin{figure*}[t]
  \centering
  \includegraphics[width=\textwidth]{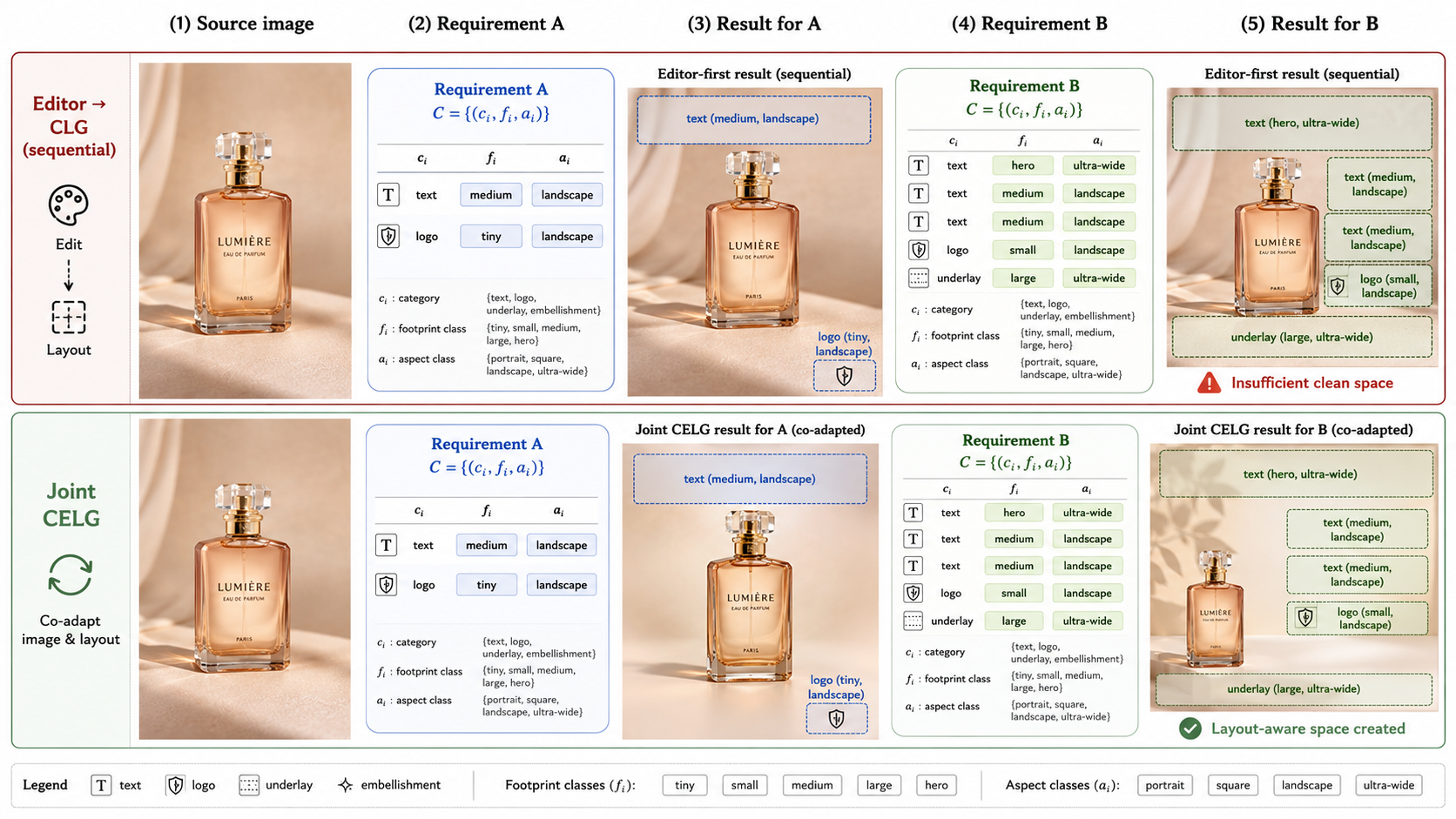}
  \caption{Limitation of sequential content-aware layout generation. Given
  the same source asset, an editor-only pipeline prepares the canvas without
  feedback from the foreground requirement. The resulting canvas accommodates
  sparse requirement A but leaves insufficient clean space for denser
  requirement B. CELG instead co-adapts the image and layout, repositioning the
  product and refining the background to create layout-aware space.}
  \label{fig:clg_limitation}
\end{figure*}

To address this limitation, we formulate \emph{content-edited layout generation} (CELG), a conditional generation task that extends CLG from prepared backgrounds to raw visual assets. Given a raw asset, a design intention, and an optional foreground requirement $\mathcal{C}$, CELG produces an edited background together with a compatible layout, covering the process of transforming an unprocessed asset into a layout-ready design template. We consider two settings: \emph{Unconstrained CELG}, where $\mathcal{C}=\varnothing$, and \emph{Requirement-Conditioned CELG}, where $\mathcal{C}\neq\varnothing$. The foreground requirement can directly affect how the canvas should be prepared: an edited background that provides sufficient space for a sparse layout may be unsuitable for a denser one. In a sequential pipeline, once the background has been edited without layout feedback, the layout generator must work within the remaining space, potentially resulting in crowding or occlusion. CELG instead allows the foreground demand to guide canvas adaptation, so that the subject and its surroundings can be adjusted before the final layout is determined. This formulation supports practical workflows that begin with unprocessed visual assets, such as transforming an ordinary product photograph into a poster-ready canvas or refining its scene, lighting, and composition while preserving the main subject.  Figure~\ref{fig:clg_limitation} illustrates this difference under sparse and dense foreground requirements.

To demonstrate why CELG should be formulated as a distinct task, we address two questions in sequence. We first ask \emph{Q1: Does transforming a raw asset into a layout-ready canvas improve the quality of the generated layout?} Q1 examines whether canvas preparation is beneficial when extending CLG to raw assets. We then ask \emph{Q2: If canvas preparation is beneficial, is a single edit before layout generation sufficient, or should the edit also be informed by layout information?} Q2 examines whether canvas preparation can be treated as an independent preprocessing step or should be coordinated with layout generation.

Answering these questions requires supervision that existing graphic-design datasets do not provide: they typically contain only a prepared canvas and its annotated layout, without the raw source asset from which that canvas was derived, and therefore lack the source--target pairs needed to train or systematically evaluate CELG. We address this gap with \emph{Search-Compare-Edit} (SCE), which reverses the usual preparation process by synthesizing a plausible pre-edit source from a prepared design canvas, yielding the \emph{InstructEdit} dataset with a source image, target canvas, editing instruction, and layout for each sample.

To tackle this task, we integrate a pretrained flow-matching image editor with a layout flow transformer using bidirectional cross-attention bridges. The two branches follow synchronized sampling trajectories and exchange information between their partially denoised states, allowing canvas adaptation and layout planning to guide each other throughout generation. Our experiments demonstrate the effectiveness of this design for CELG.

Our evaluation is organized around the two questions above. For Q1, we hold the layout generator fixed and apply it to the raw source, an automatically edited canvas, and the professionally prepared target canvas, isolating the effect of canvas readiness from that of layout-model capacity. For Q2, we compare four coordination strategies using the same editing and layout components---image-first sequential generation, layout-first refinement, edit-first iterative refinement, and joint generation---under the same source images, design intentions, and backbone access, so that their differences reflect the direction and timing of information exchange. We evaluate image quality, layout quality, Readability, Occlusion, and holistic TemplateFID on the \textit{InstructEdit} test split. For holistic evaluation, we adopt TemplateFID from InterIL~\cite{yang2026interil} and retrain its TemplateAE evaluator without any CGL samples.

Our results answer both questions. For Q1, automatically edited canvases yield substantially more compatible layouts than raw assets and approach the professional target canvases, confirming that canvas readiness can bound downstream layout quality. For Q2, Layout$\rightarrow$Edit improves every quality metric over Edit$\rightarrow$Layout at essentially the same runtime, showing the value of layout-guided canvas adaptation even in a one-way pipeline. External iteration obtains the best Occlusion and Readability but nearly doubles inference time, whereas our joint model achieves the best overall layout and holistic quality in 20.8s, providing the strongest quality--efficiency trade-off.

Our main contributions are summarized as follows:
\begin{itemize}
    \item We formulate CELG, a new task that extends layout generation from prepared canvases to raw visual assets by incorporating canvas adaptation into the generation process.
    \item We propose SCE, an automatic data construction framework that synthesizes source--target editing pairs from existing graphic designs, and use it to construct InstructEdit, a dataset of aligned source assets, target canvases, editing instructions, and layouts.
    \item We develop a two-flow model that couples a pretrained image editor with a layout flow transformer through synchronized sampling and bidirectional cross-attention, enabling canvas adaptation and layout generation to coordinate throughout the generation process. Experiments demonstrate the effectiveness of the proposed model for CELG.
\end{itemize}

\section{Related Work}
\label{sec:related_work}

\subsection{Content-aware Layout Generation}

Content-aware layout generation studies how to generate layouts that are compatible with given visual content, especially background images in graphic design. Early methods introduce visual semantics or geometric cues into layout prediction to improve compatibility between foreground elements and background content~\cite{zheng2019content, zhou2022composition, cao2022geometry}. More recent approaches have explored stronger generative backbones and richer conditioning signals. In particular, LayoutDiT~\cite{li2024layoutditexploringcontentgraphicbalance} formulates layout generation with diffusion transformers, Scan-and-Print~\cite{Hsu-IJCAI2025-ScanandPrint} improves efficiency through patch-level data summarization and vertex-based representation, and RALF~\cite{horita2024retrievalaugmented} introduces retrieval-augmented generation to exploit similar layout exemplars. Large multimodal models have also been adopted for this task, as exemplified by PosterLlama~\cite{seol2024posterllama}, which uses multimodal reasoning to generate layouts conditioned on background images.

Despite their effectiveness, these methods share a common assumption: the input image is fixed, and only the layout is generated. This assumption limits their applicability in practical design workflows, where the background image itself is often adapted to better match the design goal. Our work extends content-aware layout generation to a broader task, \textit{content-edited layout generation}, in which background editing and layout generation are performed jointly under a user-defined intention and a foreground requirement that may be empty.
More closely related to CELG, Desigen~\cite{weng2024desigen} jointly synthesizes
background templates and layouts from text. InterIL~\cite{yang2026interil}
further models interaction between pretrained image and layout priors for joint
image--layout generation and introduces TemplateFID and TemplateCLIP for
completed-template evaluation. Both methods generate a template from text
rather than adapting a user-provided image, so their inputs differ from ours
and we do not compare against them.

Beyond CLG and joint template synthesis, recent studies have explored other
aspects of graphic design generation. Order Matters~\cite{yang2025order}
investigates content-adaptive element ordering to improve generative modeling
of graphic designs. Human-aware Design Generation~\cite{hou2026humanaware}
studies a complementary design-conditioned task that completes a partial design
by generating and placing a 3D human representation. Hierarchical systems such
as COLE~\cite{jia2023cole} and IGD~\cite{qu2025igd} instead generate the textual,
graphical, and layered content of a design from a high-level brief. These works
address generation order, human-aware completion, and full design synthesis,
respectively, rather than adapting a user-provided image together with its
foreground layout.

\subsection{Instruction-Based Image Editing}

Instruction-based image editing aims to modify images according to natural language instructions. Earlier methods such as InstructPix2Pix~\cite{brooks2023instructpix2pixlearningfollowimage}, Prompt-to-Prompt~\cite{hertz2022prompt}, and Text2LIVE~\cite{bar2022text2live} demonstrate the effectiveness of language-guided image manipulation for edits such as recoloring, object replacement, and style transformation. Subsequent works improve editing quality and controllability using real-world supervision, stronger architectures, or multimodal reasoning~\cite{Zhang2023MagicBrush, huang2024smartedit, yu2024anyedit, liu2025step1x-edit}. In particular, recent flow-matching-based editors directly learn the transformation from noise to edited images under textual guidance. We use FireRed-Image-Edit-1.1~\cite{firered2026imageedit}, a general-purpose image editing model built on the Qwen-Image architecture~\cite{wu2025qwenimagetechnicalreport}, both as the editor used to construct our paired data and as the image editing backbone in our framework.

While these methods achieve strong performance on general-purpose image editing, they mainly focus on standalone visual manipulation and do not address the interaction between image editing and layout generation. In graphic design, however, background edits are often closely coupled with layout intent: edits may need to create cleaner layout space, improve readability, or better support the semantic goal of the final composition. Our work bridges these two directions by coupling a flow-matching-based image editor with a layout flow matching model, enabling joint generation of edited backgrounds and layouts for intention-driven graphic design.

\section{Method}
\label{sec:method}

\subsection{Problem Formulation}

Given a coarse source asset $I_{\text{src}}$, a natural-language design
intention $u$, and an optional foreground requirement $\mathcal{C}$, CELG maps
the three inputs to an adapted background and a foreground layout:
\begin{equation}
    (I_{\text{src}},u,\mathcal{C})
    \longmapsto
    (\tilde{I},\mathcal{L}),
    \qquad
    \mathcal{C}\in
    \left\{
        \varnothing,\,
        \{(c_i,f_i,a_i)\}_{i=1}^{M}
    \right\}.
\end{equation}
The output $\tilde{I}$ is the adapted background, and
$\mathcal{L}=\{(b_j,c_j)\}_{j=1}^{N}$ contains $N$ foreground elements,
where $b_j\in[0,1]^4$ is the normalized bounding box of element $j$ and
$c_j$ is its category. We consider two settings. In \emph{Unconstrained
CELG}, $\mathcal{C}=\varnothing$ and the model determines the number $N$ of
layout elements. In \emph{Requirement-Conditioned CELG},
$\mathcal{C}=\{(c_i,f_i,a_i)\}_{i=1}^{M}$ specifies $M$ requested elements,
and the generated layout contains one element for each request, so $N=M$.
We model at most ten foreground elements; thus, $N\leq 10$, and
$M\leq 10$ whenever $\mathcal{C}$ is nonempty.

For each requested element $i$, $c_i$ specifies its semantic category,
while $f_i$ and $a_i$ provide coarse size and shape constraints. The
\emph{footprint} class $f_i$ quantizes the normalized box area
$(x_2-x_1)(y_2-y_1)$ into five ordered levels: \textit{tiny},
\textit{small}, \textit{medium}, \textit{large}, and \textit{hero}.
Similarly, the \emph{aspect} class $a_i$ quantizes the aspect ratio
$(x_2-x_1)/(y_2-y_1)$ into four categories: \textit{portrait},
\textit{square}, \textit{landscape}, and \textit{ultra-wide}. Both labels
are computed directly from annotated
bounding boxes and therefore require no additional manual annotation.

Our approach consists of two major components: (1) a data generation method, \textit{Search-Compare-Edit} (SCE), which is used to generate synthetic paired image editing data for learning the background editing capabilities of our model; (2) a flow matching model for content-edited layout generation, which jointly generates the layout and the edited background.

\subsection{Paired Image Editing Data Generation}
\label{subsec:dataset_construction}

SCE aims to generate image editing data of the form $\{(I_\text{src}, I_\text{tgt}, \mathcal{T})\}$, where $I_\text{src}$ is the input source image, $I_\text{tgt}$ is the corresponding edited target image, and $\mathcal{T}$ is a textual instruction describing the transformation between $I_\text{src}$ and $I_\text{tgt}$.

The design of our SCE is motivated by the following important observations for background image editing in graphic design:
\begin{itemize}
    \item \textit{Source diversity}: designers draw source materials from diverse domains, including natural photographs, illustrations, and sketches. Accordingly, the source images $\{I_\text{src}\}$ should exhibit varied appearance and structural characteristics so that the trained model can adapt to real-world inputs.
    \item \textit{Limited editing degree}: once a source material is selected, designers typically make small or moderate changes rather than transforming it into an unrelated image. We operationalize this semantic proximity through a retrieval similarity threshold in the Search stage.
    \item \textit{Rich editing dimensions}: design preparation may involve changes at multiple levels: low- and mid-level appearance adjustments such as color, texture, and lighting; composition-aware spatial transformations such as repositioning or resizing salient objects to create usable layout space; and scene-level operations such as decluttering, completing or refining the scene, and adding, removing, or replacing elements. How far these edits may depart from the source depends on the application: commercial poster design typically preserves the advertised product or main subject while changing its presentation and surroundings, whereas general-purpose visual-asset creation may also modify or replace the subject itself.
\end{itemize}

\begin{figure*}[t]
  \centering
  \includegraphics[width=0.9\linewidth]{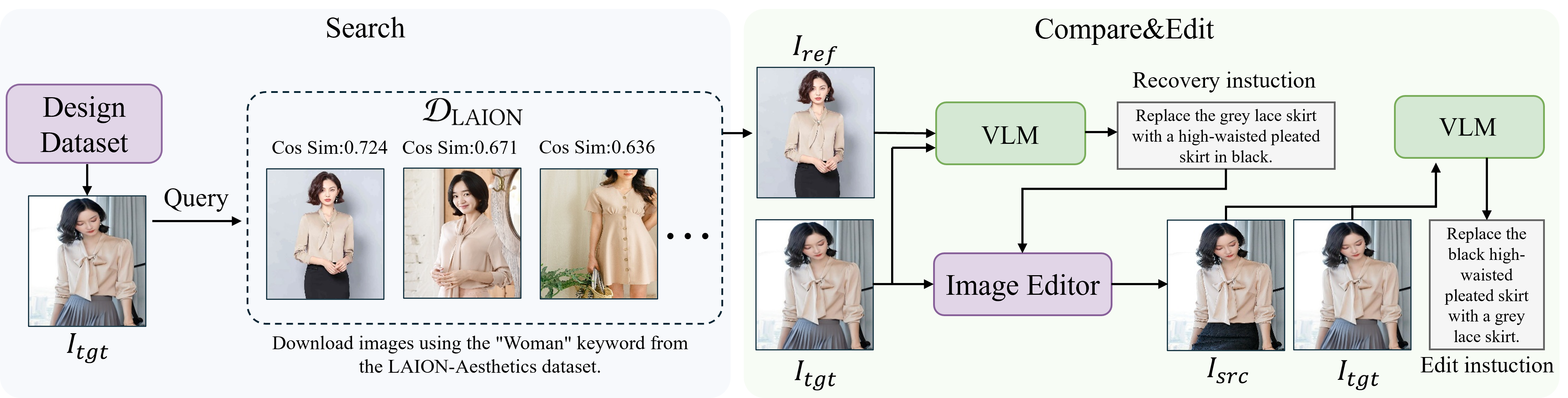}
  \caption{InstructEdit dataset construction pipeline. Semantic retrieval and multimodal distillation produce paired image-edit instructions for supervised layout-aware editing.}
  \label{fig:dataset_construction_1}
\end{figure*}

An overview of our SCE pipeline is shown in Figure~\ref{fig:dataset_construction_1}. The basic idea of SCE is to edit a target image $I_\text{tgt}$ directly taken from an existing design into a source image $I_\text{src}$ using an off-the-shelf image editing model~\cite{firered2026imageedit} based on a synthetic recovery instruction $\mathcal{T}_r$. SCE leads to an image editing dataset, \textit{InstructEdit}, which is used for supervised training of our background image editing model that will be introduced later.

\subsubsection*{Search.}
The first stage treats background images from graphic design domains as edited targets $I_\text{tgt}$ and retrieves, for each target, a \textit{source reference} $I_\text{ref}$ from a cross-domain candidate pool. The reference acts as a proxy for the source image $I_\text{src}$ and specifies the visual characteristics toward which $I_\text{tgt}$ will be reverted. Drawing candidates from multiple domains promotes source diversity, while selecting a perceptually similar reference keeps the difference between $I_\text{ref}$ and $I_\text{tgt}$ small or moderate.

We use background images from the CGL~\cite{zhou2022composition} dataset as professionally designed targets $I_{\text{tgt}}$. To obtain a visually relevant yet distinct reference image $I_{\text{ref}}$ , we adopt a class-conditional image retrieval strategy. Each $I_{\text{tgt}}$ is first assigned a high-level semantic category (e.g., person, animal, building) using a vision--language model~\cite{yang2025qwen3technicalreport}. For each category, we collect approximately 200{,}000 raw images from LAION-5B~\cite{schuhmann2022laion5bopenlargescaledataset} to form a candidate set $\mathcal{D}_{\text{LAION}}$, ensuring diverse source materials. We embed $I_{\text{tgt}}$ and all candidates into a shared feature space using DreamSim~\cite{fu2023dreamsim} and select the reference as the most similar candidate within the same semantic class:
\begin{equation}
    I_\text{ref} = \arg\max_{I \in \mathcal{D}_{\text{LAION}}}
    \cos\!\left(f(I),\, f(I_\text{tgt})\right),
\end{equation}
where $f(\cdot)$ denotes the DreamSim encoder. To ensure reliable pairing with moderate editing degree, we discard samples whose top match falls below a similarity threshold of $0.15$.

\subsubsection*{Compare.}
The second stage leverages a VLM~\cite{yang2025qwen3technicalreport} to analyze differences between $I_\text{ref}$ and $I_\text{tgt}$ across appearance, spatial composition, and scene content.  Specifically, it compares color, texture, lighting, and shape; the size and position of salient objects; background composition; and the addition, removal, or replacement of secondary objects. Based on the summarized differences, the VLM is instructed to generate a \textit{recovery instruction} $\mathcal{T}_r$ that transforms $I_\text{tgt}$ toward $I_\text{ref}$ along the relevant dimensions:
\begin{equation}
    \mathcal{T}_r = \operatorname{VLM}(I_\text{tgt}, I_\text{ref}, \mathcal{P}),
\end{equation}
where $\mathcal{P}$ is the prompt given to the VLM.

\subsubsection*{Edit.}
\begin{figure*}[t]
  \centering
  \includegraphics[width=0.95\textwidth]{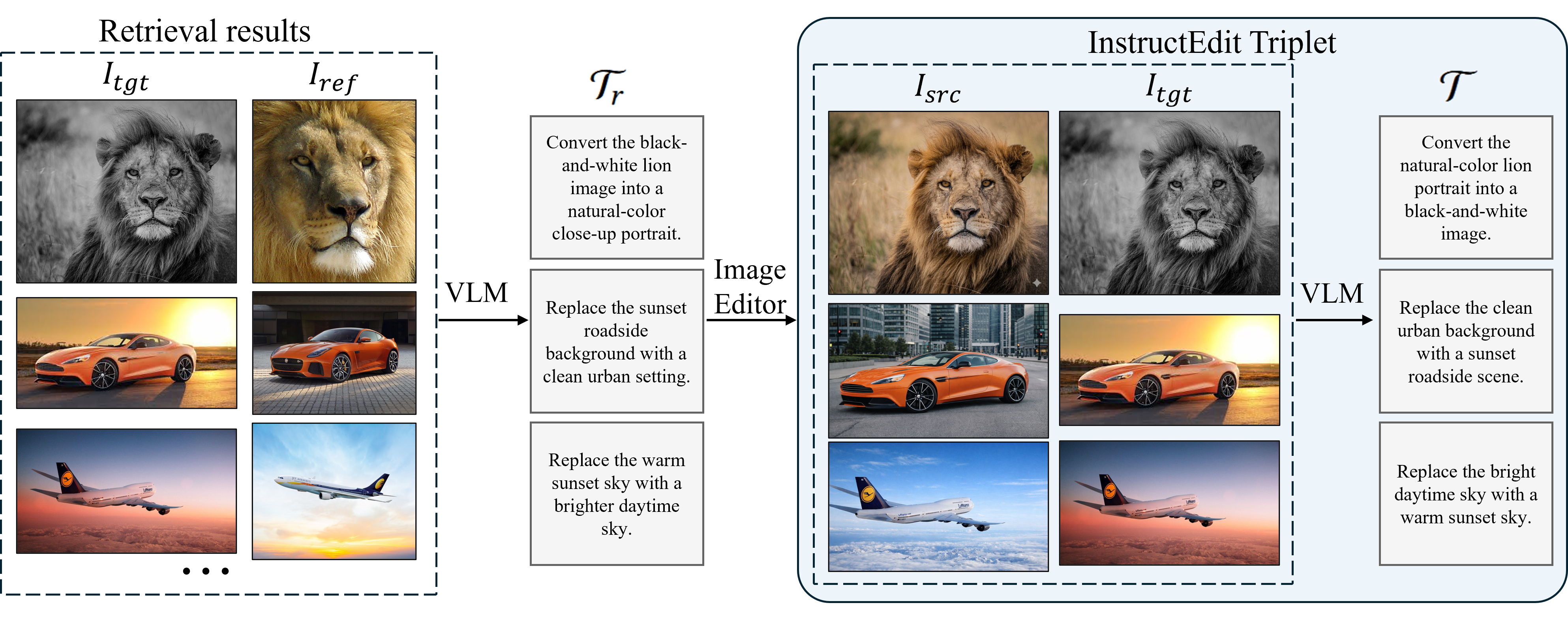}
  \caption{Examples from the SCE pipeline. For each design target
  $I_\text{tgt}$, SCE retrieves a cross-domain reference, summarizes their
  differences with a VLM, and edits the target into a source image
  $I_\text{src}$, yielding an
  $(I_\text{src}, I_\text{tgt}, \mathcal{T})$ triplet that covers appearance,
  spatial composition, and scene-level changes.}
  \label{fig:dataset_construction_2}
\end{figure*}

The third stage employs the same image editing model~\cite{firered2026imageedit} to modify $I_\text{tgt}$ based on the generated recovery instruction $\mathcal{T}_r$ to obtain the source image $I_\text{src}$:
\begin{equation}
    I_\text{src} = \operatorname{Edit}(I_\text{tgt}, \mathcal{T}_r).
\end{equation}
In this way, $I_\text{src}$ differs from $I_\text{tgt}$ along the dimensions specified by $\mathcal{T}_r$, while other essential content remains unchanged. For commercial-poster applications, preserving the advertised product relies primarily on the strong content-preserving prior acquired by the state-of-the-art image editor through large-scale pretraining, allowing its presentation and surroundings to change without altering its identity.
To obtain the final training triplet, we reapply a VLM to compare $I_{\text{src}}$ and $I_{\text{tgt}}$, generating a forward edit instruction $\mathcal{T}$ that describes the transformation from source to target:
\begin{equation}
    \mathcal{T} = \operatorname{VLM}(I_\text{src}, I_\text{tgt}, \mathcal{P}'),
\end{equation}
where $\mathcal{P}'$ is the prompt for generating the forward edit instruction.
This yields the triplet $(I_{\text{src}}, I_{\text{tgt}}, \mathcal{T})$, where
$I_{\text{src}}$ serves as the input and $I_{\text{tgt}}$ as the supervision
target during training. Representative examples are shown in
Figure~\ref{fig:dataset_construction_2}. The SCE pipeline ensures that the
generated instructions are both executable and accurately describe the
intended visual transformation.

\subsection{Content-Edited Layout Generation Model}

\begin{figure*}[t]
  \centering
  \includegraphics[width=0.95\linewidth]{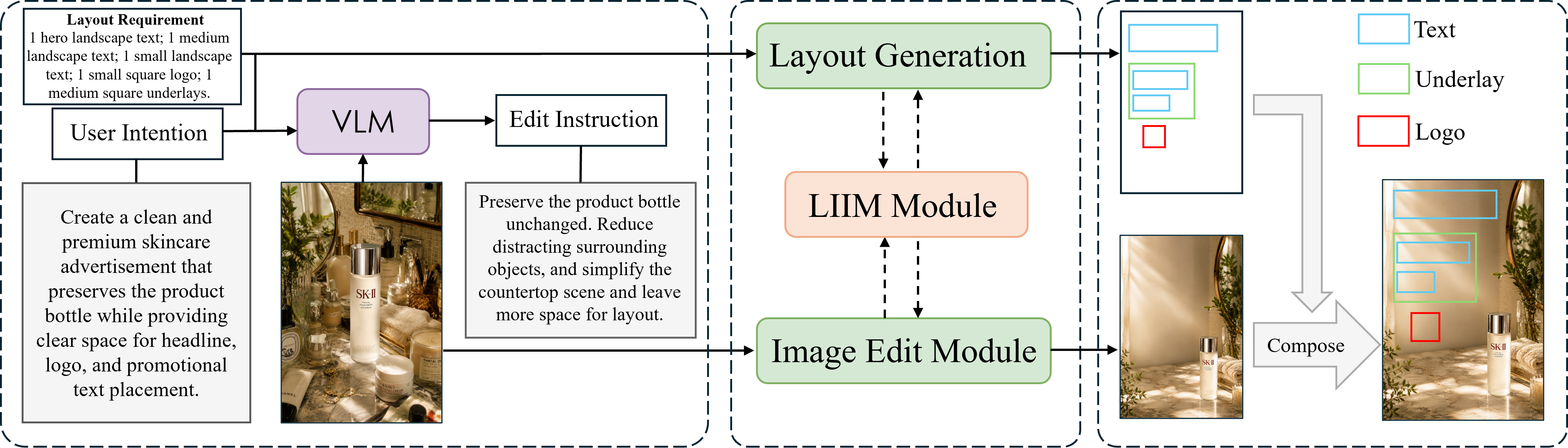}
  \caption{Overview of our content-edited layout generation framework. Given a coarse background image, a user intention, and an optional foreground requirement, we use a VLM to generate an edit instruction. This guides two coupled flow matching models---an image editing transformer and a layout DiT---connected through interleaved cross-attention bridges for bidirectional communication. }
  \label{fig:pipeline1}
\end{figure*}

Our goal is to jointly adapt the background and generate a layout that
reflects the user-provided design intention and, when specified, satisfies
the foreground requirement $\mathcal{C}$. To this end, our unified framework
performs background editing and layout generation together, with continuous
information exchange between the two branches. This bidirectional
communication allows visual content and spatial arrangement to co-evolve
throughout generation. The overall architecture is illustrated in
Figure~\ref{fig:pipeline1}.

Our framework adopts a three-stage training pipeline: (1) training a layout VAE for compact latent representation, (2) training a layout flow matching transformer (DiT) in the VAE's latent space, and (3) coupling the frozen image backbone with the mostly frozen layout backbone. In the final stage, only the communication bridges, the last four layout-DiT blocks, and its output projection are updated.

\subsubsection*{Layout VAE.}
We represent a ground-truth layout as $\mathcal{L}=\{(b_i, c_i)\}_{i=1}^{N}$, where $b_i\in[0,1]^4$ denotes normalized bounding box coordinates and $c_i$ the element category. Layouts are serialized into at most 10 element slots and padded to a fixed length. To obtain a compact continuous representation suitable for flow matching, we train a Transformer-based $\beta$-VAE~\cite{kingma2022autoencodingvariationalbayes} with $\beta=5\times10^{-4}$. The encoder $\mathcal{E}_{\text{L}}$ maps each slot to an 8-dimensional mean and log-variance, from which a latent code $l\in\mathbb{R}^{10\times8}$ is sampled via the reparameterization trick. The decoder $\mathcal{D}_{\text{L}}$ reconstructs element categories with cross-entropy and bounding boxes with coordinate regression, yielding $\hat{\mathcal{L}}=\mathcal{D}_{\text{L}}(l)$. Both encoder and decoder use 10 Transformer layers with hidden size 512 and 8 attention heads.

\subsubsection*{Layout Flow Matching Model.}
Given the pretrained VAE encoder, a ground-truth layout $\mathcal{L}$ is encoded into a clean latent $l_0 = \mathrm{Enc}(\mathcal{L})$. We then train a flow matching model in this latent space. For noise $\epsilon \sim \mathcal{N}(0, I)$, the noisy latent at timestep $\sigma = t/T \in [0, 1]$ is constructed by linear interpolation between $l_0$ and $\epsilon$:
\begin{equation}
l_\sigma = (1 - \sigma) \cdot l_0 + \sigma \cdot \epsilon.
\end{equation}
Here $\sigma{=}0$ corresponds to clean data and $\sigma{=}1$ to pure noise. The target velocity that transports $l_\sigma$ back toward $l_0$ is $v^* = \epsilon - l_0$. We train a DiT backbone~\cite{peebles2023scalable} $f_\psi$ with 28 Transformer blocks to predict this velocity $v_{\text{pred}} = f_\psi(l_\sigma, \sigma,\mathcal{C})$, minimizing:
\begin{equation}
\mathcal{L}_{\text{layout}} =
\mathbb{E}_{l_0,\, \sigma,\, \epsilon}\!
\Big[ \big\| v_{\text{pred}} - (\epsilon - l_0) \big\|_2^2 \Big].
\end{equation}
Because the VAE preserves one latent per element slot, the foreground
requirement can be injected without losing element correspondence. For slot
$i$, three independent embedding tables encode its category, footprint, and
aspect classes:
\begin{align}
e_i &= E_{\mathrm{cat}}(c_i)+E_{\mathrm{foot}}(f_i)
      +E_{\mathrm{asp}}(a_i),\\
h_i^{(0)} &= \operatorname{FC}_{\mathrm{in}}(l_{\sigma,i})+p_i+e_i,
\end{align}
where $p_i$ is a positional embedding. Thus, the timestep provides a global
condition through adaptive layer normalization, while $e_i$ remains aligned
with the element axis and is added before the first DiT block. Padded slots
use learned null entries and remain masked. During training, the complete
requirement is replaced by null entries with probability 0.1. This also
supports Unconstrained CELG and classifier-free velocity guidance at
inference:
\begin{equation}
\hat v=\hat v_{\varnothing}
 +s\bigl(\hat v_{\mathcal{C}}-\hat v_{\varnothing}\bigr),
\end{equation}
where $s$ controls the trade-off between diversity and requirement
satisfaction.

At inference, we sample pure Gaussian noise $l_{\sigma=1} \sim \mathcal{N}(0,I)$ as the starting point. The Euler sampler then iteratively integrates the predicted velocity $f_\psi(l_\sigma, \sigma,\mathcal{C})$, stepping $\sigma$ from 1 toward 0, to yield the denoised latent $l_{\sigma=0}$. The final generated layout $\hat{\mathcal{L}} = \mathrm{Dec}(l_{\sigma=0})$ is obtained by passing $l_{\sigma=0}$ through the pretrained VAE decoder.

\subsubsection*{Image Editing Module.}
For background editing, we adopt FireRed-Image-Edit-1.1~\cite{firered2026imageedit}, a pretrained flow-matching editor built on the Qwen-Image architecture~\cite{wu2025qwenimagetechnicalreport}. The same editor is used to construct \textit{InstructEdit} in Sec.~\ref{subsec:dataset_construction}. Given the source asset $I_{\text{src}}$ and an edit instruction $\mathcal{T}$, the frozen editor generates the adapted background through its original flow-matching process. During training, $\mathcal{T}$ is provided by \textit{InstructEdit}; at inference, a VLM~\cite{yang2025qwen3technicalreport} derives it from $I_{\text{src}}$, the design intention $u$, and the foreground requirement $\mathcal{C}$ when specified.

\subsubsection*{Cross-Modal Communication.}
We couple the image and layout branches through three pairs of cross-attention
bridges placed at progressively deeper layers. Before the first exchange, image
blocks 0--19 and layout blocks 0--14 operate independently to form their
initial modality-specific representations.

Each exchange round follows the same ordered procedure. Let $F_k$ and $H_k$
denote the image and layout representations entering round $k$. An
\emph{image-to-layout bridge} first updates the layout tokens using the image
tokens as context:
\begin{equation}
    \bar{H}_k
    = H_k + \mathcal{B}^{I\rightarrow L}_k(H_k,F_k),
\end{equation}
where $\mathcal{B}^{I\rightarrow L}_k$ is cross-attention with layout queries
and image keys and values. After the layout branch advances to its next
exchange point, its output $H_k^{+}$ is passed back through a
\emph{layout-to-image bridge}:
\begin{equation}
    \bar{F}_k
    = F_k + \mathcal{B}^{L\rightarrow I}_k(F_k,H_k^{+}),
\end{equation}
where image tokens serve as queries and layout tokens as keys and values. The
updated image representation then continues through the next image-transformer
segment. Learnable projections inside each bridge align the two hidden spaces.

\begin{table}[t]
\centering
\caption{Interleaved schedule of the three bidirectional exchange rounds.}
\label{tab:exchange_schedule}
\resizebox{\columnwidth}{!}{%
\begin{tabular}{cccc}
\toprule
Round & Image state supplied to layout & Layout blocks & Following image blocks \\
\midrule
1 & after blocks 0--19 & 15--19 & 20--33 \\
2 & after blocks 20--33 & 20--23 & 34--45 \\
3 & after blocks 34--45 & 24--27 & 46--59 \\
\bottomrule
\end{tabular}%
}
\end{table}

This schedule traverses all 60 image-transformer blocks and all 28 layout-DiT
blocks. The output projection of each bridge is initialized to zero so that
the coupled model initially preserves the behavior of the pretrained branches.

\subsubsection*{Training.}
We adopt a three-stage training strategy. In Stage 1, we train the layout VAE independently. In Stage 2, we train the layout DiT on the frozen VAE's latent space. In Stage 3, we freeze the pretrained image transformer and layout-DiT blocks 0--23. We optimize only the three image-to-layout bridges, the three layout-to-image bridges, layout-DiT blocks 24--27, and the output projection.

The Stage 3 training objective jointly optimizes both branches:
\begin{align}
\mathcal{L}_{\text{total}}
&= \lambda_{\text{image}}\mathcal{L}_{\mathrm{image}} + \lambda_{\text{layout}}\mathcal{L}_{\mathrm{layout}},
\end{align}
where $\lambda_{\text{image}}{=}\lambda_{\text{layout}}{=}1$. Both branches share the same timestep $t \sim \mathrm{Uniform}[0, T)$ at each training step, enabling synchronized flow matching trajectories. Through this joint optimization, the cross-attention bridges learn to exchange complementary information between modalities, producing coherent and intention-aligned design outcomes.

\section{Experiments}
\label{sec:experiment}

\subsection{Datasets}

Following LayoutDiT~\cite{li2024layoutditexploringcontentgraphicbalance}, we
use its annotated split of CGL~\cite{zhou2022composition}, comprising 48,544
training, 6,002 validation, and 6,002 test designs. Applying SCE to this split
yields \textit{InstructEdit} with 23,042 training pairs, 2,347 validation
pairs, and 6,000 test samples. All experiments are conducted on the
\textit{InstructEdit} test split.

\subsection{Implementation Details}

\noindent\textbf{Layout Branch.} We optimize the layout VAE with AdamW at a learning rate of $10^{-4}$. The 28-block layout DiT uses hidden size 512 and 8 attention heads and is trained for 1000 epochs with $T{=}1000$ flow-matching timesteps. Before DiT training, layout latents are normalized by the standard deviation measured from the first training batch. The complete foreground requirement is dropped with probability 0.1 to train the null-conditioning branch.

\noindent\textbf{Image Branch.} We use FireRed-Image-Edit-1.1~\cite{firered2026imageedit}, a pretrained flow-matching editor with 60 dual-stream blocks (hidden dimension 3072, 24 attention heads). The model operates on packed latent tokens from a VAE with 8$\times$ spatial compression and 16 latent channels.

\noindent\textbf{Communication Training.} In Stage 3, we freeze the FireRed transformer and VAE together with layout-DiT blocks 0--23. We train the three image-to-layout bridges, three layout-to-image bridges, layout-DiT blocks 24--27, and the final projection. We use AdamW with learning rate $10^{-4}$ and gradient accumulation over 4 steps.

\noindent\textbf{Inference.} We use Euler sampler with 50 steps for both layout and image generation. Classifier-free guidance is applied to the image branch with scale 4.0. Both branches share the same timestep schedule for synchronized sampling.

\subsection{Baselines}

We compare our method with four state-of-the-art content-aware layout generation methods:
\begin{itemize}
    \item \textbf{PosterLlama}~\cite{seol2024posterllama} is an LVM-based approach that performs layout generation conditioned on background images using multimodal reasoning.
    \item \textbf{LayoutDiT}~\cite{li2024layoutditexploringcontentgraphicbalance} explores the content-graphic balance in layout generation using a diffusion transformer architecture.
    \item \textbf{Scan-and-Print}~\cite{Hsu-IJCAI2025-ScanandPrint} proposes patch-level data summarization and augmentation for efficient content-aware layout generation, using a vertex-based layout representation.
    \item \textbf{RALF}~\cite{horita2024retrievalaugmented} (Retrieval-Augmented Layout Transformer) retrieves nearest neighbor layout examples based on input images and feeds them into an autoregressive generator for improved generation quality.
\end{itemize}

Although these methods are originally formulated for fixed-background layout generation, we place them in the same editor-then-layout pipeline for a fair CELG comparison and evaluate both task variants. Under \emph{Unconstrained CELG}, we set $\mathcal{C}=\varnothing$ for the four original baselines and our model. Under \emph{Requirement-Conditioned CELG}, we set $\mathcal{C}\neq\varnothing$ for our model and for the adapted versions of PosterLlama, LayoutDiT, and Scan-and-Print. These adaptations preserve the original visual backbones and generation mechanisms, so that each baseline is adapted from its released weights rather than retrained. RALF is evaluated under the unconstrained variant only, because its discrete constraint vocabulary cannot be extended without invalidating its released checkpoint; this and the three adaptations are detailed in the supplementary material.

For requirement conditioning, Scan-and-Print concatenates learned category,
footprint, and aspect tokens with its retained image patches; PosterLlama adds
the three attributes to each serialized element while leaving coordinate
fields masked; and LayoutDiT adds slot-aligned requirement embeddings to its
noisy element tokens while clamping the supplied category channels. All
methods use the same dataset-level quantile bins for footprint and aspect.

Under Unconstrained CELG, the common VLM derives the edit instruction from
$I_{\mathrm{src}}$ and $u$. Under Requirement-Conditioned CELG, it additionally
receives a textual description of the requested element categories,
footprints, and aspect classes in $\mathcal{C}$ and produces a common
$\mathcal{C}$-aware instruction $\mathcal{T}_{\mathcal{C}}$. The exact same
instruction conditions the shared pretrained image editor for every baseline
and the image branch of our model. Each CLG baseline then generates its layout
from the edited canvas. In Panel B of Table~\ref{tab:baseline_metrics}, the
layout generator additionally receives $\mathcal{C}$. Thus, methods within
each panel of Table~\ref{tab:baseline_metrics} receive the same inputs, edit instruction, and editing backbone;
the baselines differ by organizing editing and layout generation sequentially
without layout-to-image feedback. All methods are evaluated on the
\textit{InstructEdit} test split.

\subsection{Evaluation Metrics}

We evaluate our framework from both \textbf{background image quality} and \textbf{layout quality} perspectives using a comprehensive set of quantitative metrics. Let $f_{\text{img}}$ denote a CLIP image encoder and $\mathcal{E}_{\text{L}}$ a learned layout encoder.

\noindent\textbf{Background Image Quality.}
We report the \textit{CLIP Score}~\cite{radford2021learning} between the output background and the design intention.

\noindent\textbf{Layout Evaluation.}
We evaluate both requirement satisfaction and the overall geometric quality of
the generated layouts. In the requirement-conditioned setting, all methods are
given the same foreground requirement $\mathcal{C}$, allowing us to compare how
well their outputs accommodate the requested content. We measure this using
\textit{Requirement Satisfaction} ($\mathcal{C}$-Sat.), which considers the
category, footprint, and aspect class of each element. Each generated box is
first assigned a footprint and aspect class using the same thresholds employed
to construct $\mathcal{C}$. A requested element can then be matched to a
generated element only when all three attributes agree, with each element used
in at most one match. Let $m^\star$ denote the largest possible number of such
matches. For $M$ requested elements and $N$ generated elements,
\begin{equation}
    \operatorname{C\text{-}Sat}=\frac{2m^\star}{M+N}.
\end{equation}
The score is independent of element order and decreases when requested elements
are missing or additional elements are generated. In our conditioned setting,
$N=M$, so the score simplifies to $m^\star/M$. For general layout quality,
\textit{Alignment}~\cite{li2019layoutgan} evaluates the geometric regularity and
positional consistency among elements, while
\textit{Overlay}~\cite{Hsu-2023-posterlayout} measures their average pairwise
IoU, excluding underlay elements, to quantify element collisions.
\textit{Layout Fr\'echet Distance (L-FID)}~\cite{Kikuchi2021} compares the
distributions of generated and real layouts in the representation space of the
frozen layout VAE used by TemplateAE. This VAE is trained on the CGL-held-out
four-source corpus described below and shares no weights with any evaluated
generator.

\noindent\textbf{Holistic Image--Layout Evaluation.}
Beyond image-only and layout-only metrics, we adopt metrics that require joint consideration of both image and layout. \textit{Occlusion}~\cite{cao2022geometry} penalizes cases where layout elements obscure salient regions in the background, and \textit{Readability}~\cite{horita2024retrievalaugmented} quantifies the visual contrast between text elements and the background, reflecting the legibility of text-based components.

We adopt TemplateFID and TemplateCLIP from InterIL~\cite{yang2026interil} to
evaluate the generated background and layout jointly. To prevent evaluator
training from overlapping with the target CGL benchmark~\cite{zhou2022composition},
we retrain the \textit{Layout--Image Autoencoder} (TemplateAE) on
PKU~\cite{Hsu-2023-posterlayout}, WebDesign~\cite{weng2024desigen},
GenPoster-100K~\cite{wang2025sega}, and
Crello~\cite{yamaguchi2021canvasvae}, excluding all CGL samples. After
remapping annotations to a shared four-class taxonomy and discarding layouts
with more than 10 elements, the corpus contains 148,529 training templates
and 19,397 validation templates. The resulting evaluator is trained independently and
shares no weights with the CELG generator. \textit{TemplateFID} measures the
Fr\'echet distance between the distributions of generated and real
background--layout pairs in the learned joint representation space, capturing
their overall visual and structural similarity. \textit{TemplateCLIP} measures
the semantic alignment between a generated background--layout pair and the
input design intention. Full evaluator training details are provided in the
supplementary material.

\begin{table}[t]
\centering
\caption{Effect of canvas preparation on layout generation with a fixed LayoutDiT.}
\label{tab:canvas_readiness}
\resizebox{\columnwidth}{!}{
\begin{tabular}{lcccccc}
\toprule
& \multicolumn{3}{c}{Layout}
& \multicolumn{3}{c}{Image--Layout Compatibility} \\
\cmidrule(lr){2-4}
\cmidrule(lr){5-7}
Canvas Input
& L-FID$\downarrow$
& Align.$\downarrow$
& Ove.$\downarrow$
& Occ.$\downarrow$
& Read.$\downarrow$
& TemplateFID$\downarrow$ \\
\midrule
Raw source
& 0.45 & 0.014 & 0.004 & 18.73\% & 8.26\% & 79.54 \\
Edited canvas
& \underline{0.42} & \underline{0.009} & \textbf{0.001} & \underline{12.18\%} & \underline{1.63\%} & \underline{68.01} \\
Target (oracle)
& \textbf{0.40} & \textbf{0.008} & \textbf{0.001} & \textbf{9.07\%} & \textbf{0.94\%} & \textbf{67.43} \\
\bottomrule
\end{tabular}
}
\end{table}

\subsection{Q1: Does Canvas Preparation Improve Layout Quality?}

To answer Q1, we hold the layout generator fixed and vary only its input
canvas. For each test sample, we apply the same LayoutDiT to the raw source
$I_{\text{src}}$, an automatically edited canvas produced without layout
feedback, and the professionally prepared target $I_{\text{tgt}}$, which
serves as an oracle canvas. All three conditions use identical sampling
settings and omit $\mathcal{C}$, so their differences isolate the effect of
canvas preparation. As shown in
Table~\ref{tab:canvas_readiness}, automatic editing reduces Occlusion from
18.73\% to 12.18\%, Readability error from 8.26\% to 1.63\%, and TemplateFID from
79.54 to 68.01. The oracle further lowers Occlusion to 9.07\% but only
marginally improves TemplateFID to 67.43, confirming that canvas readiness
constrains downstream layout quality and that automatic preparation captures
most of the attainable compatibility gain. These results answer Q1 by showing
that canvas preparation is an important prerequisite for layout generation
from raw assets. Whether it should be performed independently of layout
planning is addressed next in Q2.

\subsection{Q2: Should Canvas Preparation Be Informed by Layout Feedback?}

To answer Q2, we evaluate layout feedback from two complementary perspectives.
We first compare our method with existing CLG baselines in
Table~\ref{tab:baseline_metrics}, where each baseline is placed in the same
editor-then-layout CELG pipeline. We then compare alternative editing--layout
coordination strategies using the same model components, isolating whether and
when layout information should guide canvas adaptation.

\subsubsection{Comparison with Existing Methods}

\begin{table*}[t]
\centering
\caption{Comparison under Unconstrained and Requirement-Conditioned CELG.
In Panel B, the same nonempty $\mathcal{C}$ conditions both the common edit
instruction and each layout generator.}
\label{tab:baseline_metrics}
\resizebox{0.95\textwidth}{!}{
\begin{tabular}{lccccccccc}
\toprule
& \multicolumn{1}{c}{Image}
& \multicolumn{4}{c}{Layout}
& \multicolumn{4}{c}{Holistic (Image + Layout)} \\
\cmidrule(lr){2-2}
\cmidrule(lr){3-6}
\cmidrule(lr){7-10}
Method &
CLIP$\uparrow$ &
$\mathcal{C}$-Sat.$\uparrow$ &
L-FID$\downarrow$ &
Align.$\downarrow$ &
Ove$\downarrow$ &
Occ.$\downarrow$ &
Read.$\downarrow$ &
TemplateFID$\downarrow$ &
TemplateCLIP$\uparrow$ \\
\midrule

\multicolumn{10}{l}{\textbf{Panel A: Unconstrained CELG}} \\

PosterLlama
& 29.84 & -- & 0.63 & 0.024 & 0.031
& 15.93\% & 10.54\% & 74.65 & 2.89 \\

Scan-and-Print
& 29.84 & -- & 1.44 & 0.034 & 0.023
& \textbf{6.42\%} & 2.42\% & 79.28 & 2.83 \\

RALF
& 29.84 & -- & 1.32 & \underline{0.008} & 0.042
& 19.13\% & 2.93\% & 77.48 & 2.79 \\

LayoutDiT
& 29.84 & -- & \underline{0.42} & 0.009
& \textbf{0.001} & 12.18\% & \underline{1.63\%}
& \underline{68.01} & \underline{3.04} \\

Ours (unconstrained)
& \textbf{30.87} & -- & \textbf{0.38} & \textbf{0.006}
& \textbf{0.001} & \underline{11.38\%} & \textbf{1.48\%}
& \textbf{66.24} & \textbf{3.18} \\

\midrule
\multicolumn{10}{l}{\textbf{Panel B: Requirement-Conditioned CELG}} \\

PosterLlama ($+\mathcal{C}$)
& 29.84 & 68.7\% & 0.57 & 0.022 & 0.027
& 14.21\% & 8.36\% & 72.18 & 2.96 \\

Scan-and-Print ($+\mathcal{C}$)
& 29.84 & \underline{76.2\%} & 1.31 & 0.031 & 0.019
& \textbf{6.81\%} & 2.03\% & 76.54 & 2.91 \\

LayoutDiT ($+\mathcal{C}$)
& 29.84 & 72.1\% & \underline{0.40}
& \underline{0.009} & \underline{0.002}
& 10.47\% & \underline{1.41\%}
& \underline{66.11} & \underline{3.08} \\

Ours ($+\mathcal{C}$)
& \textbf{30.95} & \textbf{82.4\%} & \textbf{0.35}
& \textbf{0.006} & \textbf{0.001}
& \underline{7.24\%} & \textbf{1.28\%}
& \textbf{62.73} & \textbf{3.24} \\

\bottomrule
\end{tabular}
}
\end{table*}

\paragraph{Quantitative Results}

Table~\ref{tab:baseline_metrics} presents the comparison on the
\textit{InstructEdit} test set under the two task variants. Panel A of
Table~\ref{tab:baseline_metrics} evaluates Unconstrained CELG with
$\mathcal{C}=\varnothing$. Panel B of the same table supplies the same nonempty
$\mathcal{C}$ to the common VLM prompt and every layout generator;
all methods therefore use the exact same $\mathcal{C}$-aware edit instruction
and pretrained image editor. The CLG baselines generate their layouts after
the image edit is fixed, whereas our method jointly generates the edited canvas
and layout, allowing intermediate layout predictions to guide canvas adaptation
and image features to guide element placement throughout sampling.

\noindent\textbf{Unconstrained CELG.}
The four editor-then-layout baselines share the same edited background and
therefore the same CLIP score of 29.84, whereas our method reaches 30.87.
It also achieves the best L-FID (0.38), Alignment (0.006), Readability
(1.48\%), TemplateFID (66.24), and TemplateCLIP (3.18), while tying LayoutDiT on
Overlay (0.001). Scan-and-Print retains the lowest Occlusion, but its poorer
layout and holistic scores show that avoiding salient regions alone does not
ensure a coherent completed design. Overall, these results show that jointly
generating the edited canvas and layout achieves better image, layout, and
holistic quality than sequential pipelines that first edit the image and then
generate the layout. Additional qualitative comparisons for Unconstrained
CELG are provided in the supplementary material.

\noindent\textbf{Requirement-Conditioned CELG.}
Our model obtains the highest C-Sat.\ at 82.4\%, exceeding the strongest
baseline, Scan-and-Print, by 6.2 percentage points, showing that its gains are
not achieved by omitting difficult required elements.
It simultaneously achieves the best CLIP score (30.95), L-FID (0.35),
Alignment (0.006), Overlay (0.001), Readability (1.28\%), TemplateFID (62.73),
and TemplateCLIP (3.24).
Relative to LayoutDiT, the strongest sequential baseline on the holistic
metrics, our method lowers TemplateFID by 3.38 and raises TemplateCLIP by 0.16.
Scan-and-Print remains slightly better on Occlusion (6.81\% versus 7.24\%),
but is substantially worse on layout realism and holistic quality. These
results indicate that jointly adapting the canvas and layout better
accommodates the requested foreground without sacrificing overall coherence.

\paragraph{Qualitative Results}

\begin{figure*}[t]
  \centering
  {\scriptsize
  \setlength{\tabcolsep}{0pt}
  \begin{tabular}{@{}
    p{0.1424\textwidth}p{0.1350\textwidth}p{0.1373\textwidth}%
    p{0.1477\textwidth}p{0.1477\textwidth}p{0.1471\textwidth}%
    p{0.1428\textwidth}@{}}
  \centering Source Asset &
  \centering Design Intention &
  \centering Requirement $\mathcal{C}$ &
  \centering PosterLlama &
  \centering Scan-and-Print &
  \centering LayoutDiT &
  \centering Ours
  \end{tabular}}
  \vspace{2pt}
  \includegraphics[width=\textwidth,trim=0 0 0 22,clip]{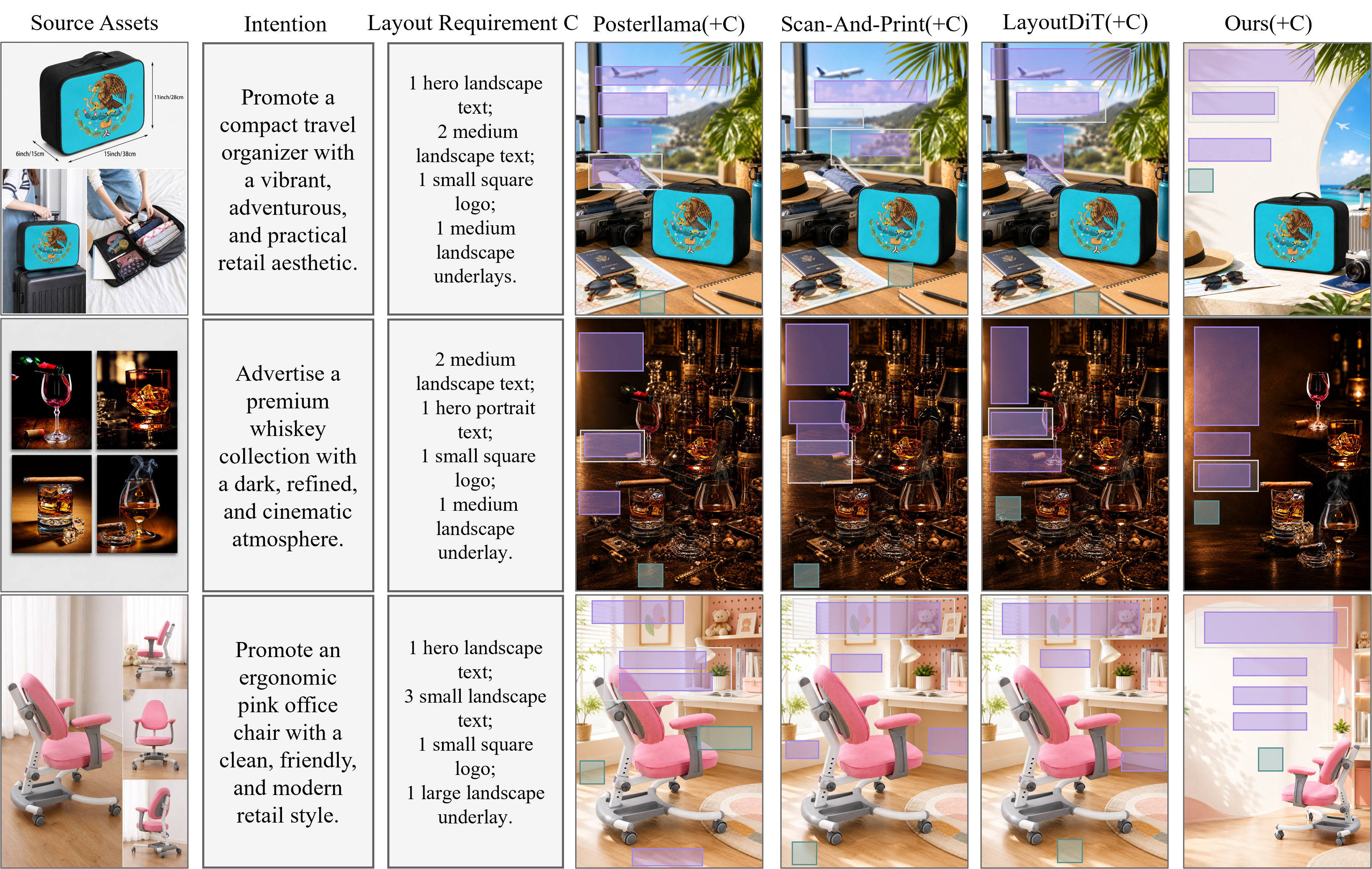}
  \caption{Requirement-conditioned qualitative comparison corresponding to
  Panel B of Table~\ref{tab:baseline_metrics}. Each row shows a source asset,
  design intention, foreground requirement, and the results of three
  sequential CLG baselines and Ours. All methods receive the same
  $\mathcal{C}$-aware edit instruction; the baselines share its edited canvas,
  whereas Ours can continue adapting the canvas through image--layout
  communication. The complete edit instructions are provided in the
  supplementary material and omitted here for space. Purple boxes denote
  text, white boxes denote underlays, and cyan boxes denote logos.}
  \label{fig:qualitative_results}
\end{figure*}

Figure~\ref{fig:qualitative_results} illustrates the Requirement-Conditioned
comparison reported in Panel B of Table~\ref{tab:baseline_metrics}. Although the sequential baselines receive both
$\mathcal{C}$ and the same $\mathcal{C}$-aware edit instruction, their canvas
is fixed before layout generation. Consequently, the requested foreground
elements may still compete with the product or visually salient background
regions. Ours uses the same instruction but allows the canvas to respond to
the evolving layout, repositioning or resizing the preserved product and
creating cleaner foreground space. Across the travel organizer, whisky, and
office-chair cases, this produces layouts that better accommodate the required
element footprints and form more coherent completed designs.

\subsubsection{Editing--Layout Coordination Strategies}

To isolate the effect of layout feedback, Table~\ref{tab:organization_ablation}
compares five Requirement-Conditioned CELG strategies on the complete
\textit{InstructEdit} test split. All strategies use the same
FireRed-Image-Edit-1.1 editor, requirement-conditioned layout flow DiT,
Qwen3-VL, inputs, and sampling configuration, but differ in the order and
information flow between editing and layout generation.
\textit{Edit$\rightarrow$Layout} follows the protocol used in
Panel B of Table~\ref{tab:baseline_metrics}: $\mathcal{C}$ conditions both the VLM-generated edit
instruction and the layout generator.
\textit{Edit$\rightarrow$Layout (w/o $\mathcal{C}$ prompt)} removes
$\mathcal{C}$ only from the VLM prompt while retaining it in the layout
generator, serving as an information ablation.
\textit{Layout$\rightarrow$Edit} instead generates and fixes a layout from
$I_{\mathrm{src}}$ and $\mathcal{C}$ with image-to-layout communication
enabled, after which the VLM derives a layout-aware instruction from
$I_{\mathrm{src}}$, the rendered layout, $u$, and $\mathcal{C}$ for adapting
the source.
\textit{Iterative} starts from the standard $\mathcal{C}$-aware
Edit$\rightarrow$Layout result, critiques it with the VLM under the same
$\mathcal{C}$, then re-edits the original source and regenerates the layout,
whereas
\textit{Ours (Joint)} exchanges partially denoised image and layout states
through bidirectional bridges along one synchronized trajectory. The two
edit-first variants and Layout$\rightarrow$Edit invoke the editor, layout
generator, and VLM once, while
Iterative invokes each twice. We report the layout and holistic metrics from
Table~\ref{tab:baseline_metrics} together with end-to-end runtime, including
all editing, layout-generation, and VLM calls; image-only CLIP is omitted
because the experiment focuses on image--layout interaction.

\begin{table*}[t]
\centering
\caption{Controlled comparison of editing--layout coordination strategies.
Time is the average runtime per sample on a single NVIDIA H200 GPU at
$512{\times}768$ resolution and includes all generation and VLM calls.}
\label{tab:organization_ablation}
\setlength{\tabcolsep}{2.0pt}
\resizebox{\textwidth}{!}{
\begin{tabular}{lccccccccc}
\toprule
& \multicolumn{4}{c}{Layout}
& \multicolumn{4}{c}{Holistic (Image + Layout)}
& \multicolumn{1}{c}{Efficiency} \\
\cmidrule(lr){2-5}
\cmidrule(lr){6-9}
\cmidrule(lr){10-10}
Method &
$\mathcal{C}$-Sat.$\uparrow$ &
L-FID$\downarrow$ &
Align.$\downarrow$ &
Ove.$\downarrow$ &
Occ.$\downarrow$ &
Read.$\downarrow$ &
TemplateFID$\downarrow$ &
TemplateCLIP$\uparrow$ &
Time$\downarrow$ \\
\midrule

Edit$\rightarrow$Layout (w/o $\mathcal{C}$ prompt)
& 76.4\%
& 0.44
& 0.012
& 0.008
& 11.32\%
& 2.86\%
& 67.48
& 3.12
& \underline{19.7s} \\

Edit$\rightarrow$Layout
& 76.9\%
& 0.42
& 0.011
& 0.007
& 10.89\%
& 2.79\%
& 66.71
& 3.14
& 19.9s \\

Layout$\rightarrow$Edit
& 78.9\%
& 0.37
& 0.009
& 0.003
& 9.61\%
& 2.74\%
& 65.92
& 3.17
& \textbf{19.6s} \\

Iterative
& \underline{81.6\%}
& \underline{0.36}
& \underline{0.007}
& \textbf{0.001}
& \textbf{6.88\%}
& \textbf{1.19\%}
& \underline{63.41}
& \underline{3.22}
& 39.4s \\

Ours (Joint)
& \textbf{82.4\%}
& \textbf{0.35}
& \textbf{0.006}
& \textbf{0.001}
& \underline{7.24\%}
& \underline{1.28\%}
& \textbf{62.73}
& \textbf{3.24}
& 20.8s \\

\bottomrule
\end{tabular}
}
\end{table*}

Table~\ref{tab:organization_ablation} separates the benefit of feedback from
its implementation cost. Removing $\mathcal{C}$ from the edit prompt degrades
every quality metric relative to the standard Edit$\rightarrow$Layout
pipeline, lowering $\mathcal{C}$-Sat.\ from 76.9\% to 76.4\% and increasing
TemplateFID from 66.71 to 67.48. Layout$\rightarrow$Edit, which supplies a realized
layout rather than only coarse requirements, further improves every quality
metric at a comparable runtime, raising $\mathcal{C}$-Sat.\ to 78.9\% and
lowering TemplateFID to 65.92. Iterative refinement achieves the best Occlusion (6.88\%) and
Readability (1.19\%) and ties the best Overlay (0.001), but doubles latency to
39.4s. Ours (Joint) achieves the best $\mathcal{C}$-Sat.\ (82.4\%), L-FID
(0.35), Alignment (0.006), TemplateFID (62.73), and TemplateCLIP (3.24), while tying the
best Overlay and remaining close to Iterative on Occlusion and Readability. It
requires 20.8s, reducing latency by 47\% relative to Iterative and thereby
providing the strongest overall quality--efficiency trade-off.

Table~\ref{tab:communication_ablation} further isolates the contribution of
each communication direction while keeping the two backbones and sampling
configuration fixed. Both one-way variants outperform independent sampling:
image-to-layout communication reduces Occlusion from 16.72\% to 11.62\%, while
layout-to-image communication raises $\mathcal{C}$-Sat.\ to 79.1\% and lowers
Readability error to 1.47\%. Bidirectional communication performs best on all
reported metrics, reaching 82.4\% $\mathcal{C}$-Sat.\ and 62.73 TemplateFID.
Together with Table~\ref{tab:organization_ablation}, these results answer Q2:
layout feedback improves canvas preparation, and bidirectional coordination
captures this benefit without the additional inference cost of external
iteration. Taken together, Q1 and Q2 show why CELG is needed: extending layout
generation to raw assets requires both canvas preparation and coordination
between that preparation and layout planning, neither of which is modeled by
conventional CLG.

\begin{table}[t]
\centering
\caption{Ablation of cross-modal communication directions.}
\label{tab:communication_ablation}
\setlength{\tabcolsep}{3.0pt}
\resizebox{\columnwidth}{!}{
\begin{tabular}{lccccc}
\toprule
Variant &
$\mathcal{C}$-Sat.$\uparrow$ &
L-FID$\downarrow$ &
Occ.$\downarrow$ &
Read.$\downarrow$ &
TemplateFID$\downarrow$ \\
\midrule

No communication
& 76.1\%
& 0.43
& 16.72\%
& 2.95\%
& 77.61 \\

image$\rightarrow$layout
& 78.3\%
& 0.40
& \underline{11.62\%}
& 1.78\%
& \underline{68.38} \\

layout$\rightarrow$image
& \underline{79.1\%}
& \underline{0.38}
& 11.91\%
& \underline{1.47\%}
& 69.16 \\

Bidirectional (Ours)
& \textbf{82.4\%}
& \textbf{0.35}
& \textbf{7.24\%}
& \textbf{1.28\%}
& \textbf{62.73} \\

\bottomrule
\end{tabular}
}
\end{table}

\subsection{Human Evaluation}

We conduct an anonymous two-stage user study with 34 participants on 30
requirement-conditioned cases. Each participant evaluates seven cases against
each of the three CLG baselines, plus one excluded quality-control trial,
yielding 238 comparisons per baseline. Methods are anonymized, left--right
order is randomized, and source cases are not repeated for a participant.
Participants first rate the edited canvases for edit quality, product
preservation, and canvas utility, and then compare the resulting templates in
terms of layout quality and holistic quality. Because all sequential baselines share the same
edited canvas, Panel A of Table~\ref{tab:human_user_study} pools its ratings as
``Sequential Editing.''

\begin{table}[t]
\centering
\caption{Human evaluation of (a) edited canvases using five-point ratings and
(b) templates using Win/Tie/Loss percentages for Ours.}
\label{tab:human_user_study}
\begin{minipage}[t]{0.49\columnwidth}
\centering
\textit{(a) Edited Canvas Ratings}\par\vspace{1pt}
\renewcommand{\arraystretch}{1.16}
\resizebox{\linewidth}{!}{
\begin{tabular}{lccc}
\toprule
Method &
Edit$\uparrow$ &
Pres.$\uparrow$ &
Utility$\uparrow$ \\
\midrule
Sequential
& 3.71
& 3.86
& 3.42 \\

Ours
& \textbf{4.08}
& \textbf{4.12}
& \textbf{4.27} \\
\bottomrule
\end{tabular}
}
\end{minipage}\hfill
\begin{minipage}[t]{0.49\columnwidth}
\centering
\textit{(b) Templates}\par\vspace{1pt}
\resizebox{\linewidth}{!}{
\begin{tabular}{lcc}
\toprule
Comparator &
Layout &
Holistic \\
\midrule

LayoutDiT ($+\mathcal{C}$)
& 59/17/24
& 63/15/22 \\

Scan-and-Print ($+\mathcal{C}$)
& 66/14/20
& 68/13/19 \\

PosterLlama ($+\mathcal{C}$)
& 71/12/17
& 73/11/16 \\

\bottomrule
\end{tabular}
}
\end{minipage}
\end{table}

Compared with sequential editing, our method improves edit quality,
preservation, and canvas utility by 0.37, 0.26, and 0.85 points, respectively.
It is also preferred over every baseline for the resulting templates, with win rates of
59--71\% on layout quality and 63--73\% on holistic quality. The especially
large utility gain indicates that layout feedback creates more usable
foreground space and benefits the final composition.

\section{Conclusion}
\label{sec:conclusion}

We introduced \textit{content-edited layout generation} (CELG), extending CLG
from prepared canvases to raw assets by generating an adapted canvas and a
compatible layout from a source asset, design intention, and optional
foreground requirement. \textit{Search-Compare-Edit} reverse-constructs aligned
raw-to-prepared image pairs to produce the \textit{InstructEdit} benchmark. Our
two-flow method couples pretrained image editing and layout generation through
synchronized trajectories and bidirectional cross-attention. Experiments in
both task settings systematically validate CELG and demonstrate the performance
advantages of our method.

\bibliographystyle{IEEEtran}
\bibliography{main}

@IEEEtranBSTCTL{BSTcontrol,
  CTLuse_url = "no",
  CTLuse_forced_etal = "yes",
  CTLmax_names_forced_etal = "6",
  CTLnames_show_etal = "3"
}

@String(CVPR  = {IEEE Conf. Comput. Vis. Pattern Recog.})

@String(ICCV  = {Int. Conf. Comput. Vis.})

@String(ICLR  = {Int. Conf. Learn. Represent.})

@String(TOG   = {ACM Trans. Graph.})

@String(CVPR  = {CVPR})

@String(ICCV  = {ICCV})

@String(ICLR  = {ICLR})

@String(TOG   = {ACM TOG})

@misc{weng2024desigen,
      title={Desigen: A Pipeline for Controllable Design Template Generation},
      author={Haohan Weng and Danqing Huang and Yu Qiao and Zheng Hu and Chin-Yew Lin and Tong Zhang and C. L. Philip Chen},
      year={2024},
      eprint={2403.09093},
      archivePrefix={arXiv},
      primaryClass={cs.CV}
}

@inproceedings{zhou2022composition,
  title     = {Composition-aware Graphic Layout GAN for Visual-Textual Presentation Designs},
  author    = {Zhou, Min and Xu, Chenchen and Ma, Ye and Ge, Tiezheng and Jiang, Yuning and Xu, Weiwei},
  booktitle = {Proceedings of the Thirty-First International Joint Conference on Artificial Intelligence, {IJCAI-22}},
  publisher = {International Joint Conferences on Artificial Intelligence Organization},
  editor    = {Lud De Raedt},
  pages     = {4995--5001},
  year      = {2022},
  month     = {7},
  note      = {AI and Arts},
  doi       = {10.24963/ijcai.2022/692},
  url       = {https://doi.org/10.24963/ijcai.2022/692},
}

@article{zheng2019content,
  title={Content-aware generative modeling of graphic design layouts},
  author={Zheng, Xinru and Qiao, Xiaotian and Cao, Ying and Lau, Rynson WH},
  journal={ACM Transactions on Graphics (TOG)},
  volume={38},
  number={4},
  pages={1--15},
  year={2019},
  publisher={ACM New York, NY, USA}
}

@inproceedings{cao2022geometry,
  title={Geometry aligned variational transformer for image-conditioned layout generation},
  author={Cao, Yunning and Ma, Ye and Zhou, Min and Liu, Chuanbin and Xie, Hongtao and Ge, Tiezheng and Jiang, Yuning},
  booktitle={Proceedings of the 30th ACM International Conference on Multimedia},
  pages={1561--1571},
  year={2022}
}

@inproceedings{yamaguchi2021canvasvae,
  title={Canvasvae: Learning to generate vector graphic documents},
  author={Yamaguchi, Kota},
  booktitle={Proc. IEEE/CVF Int. Conf. Comput. Vis. (ICCV)},
  pages={5481--5489},
  year={2021}
}

@article{jia2023cole,
  title={Cole: A hierarchical generation framework for graphic design},
  author={Jia, Peidong and Li, Chenxuan and Liu, Zeyu and Shen, Yichao and Chen, Xingru and Yuan, Yuhui and Zheng, Yinglin and Chen, Dong and Li, Ji and Xie, Xiaodong and others},
  journal={arXiv preprint arXiv:2311.16974},
  year={2023}
}

@article{seol2024posterllama,
  title   = {PosterLlama: Bridging Design Ability of Language Models to Content-Aware Layout Generation},
  author  = {Seol, Jaejung and Kim, Seojun and Yoo, Jaejun},
  journal = {arXiv preprint arXiv:2404.00995},
  year    = {2024}
}

@inproceedings{inoue2023layoutdm,
  title={LayoutDM: Discrete Diffusion Model for Controllable Layout Generation},
  author={Inoue, Naoto and Kikuchi, Kotaro and Simo-Serra, Edgar and Otani, Mayu and Yamaguchi, Kota},
  booktitle={Proceedings of the IEEE/CVF Conference on Computer Vision and Pattern Recognition},
  pages={10167--10176},
  year={2023}
}

@inproceedings{Kikuchi2021,
    title = {Constrained Graphic Layout Generation via Latent Optimization},
    author = {Kotaro Kikuchi and Edgar Simo-Serra and Mayu Otani and Kota Yamaguchi},
    booktitle = {ACM International Conference on Multimedia},
    series = {MM '21},
    year = {2021},
    pages = {88--96},
    doi = {10.1145/3474085.3475497}
}

@inproceedings{li2019layoutgan,
  author       = {Jianan Li and
                  Jimei Yang and
                  Aaron Hertzmann and
                  Jianming Zhang and
                  Tingfa Xu},
  title        = {LayoutGAN: Generating Graphic Layouts with Wireframe Discriminators},
  booktitle    = {Proc. Int. Conf. Learn. Represent. (ICLR)},
  year         = {2019},
}

@inproceedings{fu2023dreamsim,
title={DreamSim: Learning New Dimensions of Human Visual Similarity using Synthetic Data},
author= {Fu, Stephanie and Tamir, Netanel and Sundaram, Shobhita and Chai, Lucy and Zhang, Richard and Dekel, Tali and Isola, Phillip},
booktitle={Advances in Neural Information Processing Systems},
pages={50742--50768},
volume={36},
year={2023}
}

@inproceedings{radford2021learning,
  title={Learning transferable visual models from natural language supervision},
  author={Radford, Alec and Kim, Jong Wook and Hallacy, Chris and Ramesh, Aditya and Goh, Gabriel and Agarwal, Sandhini and Sastry, Girish and Askell, Amanda and Mishkin, Pamela and Clark, Jack and others},
  booktitle={International conference on machine learning},
  pages={8748--8763},
  year={2021},
  organization={PMLR}
}

@inproceedings{qu2025igd,
  title={{IGD}: Instructional Graphic Design with Multimodal Layer Generation},
  author={Qu, Yadong and others},
  booktitle={Proceedings of the IEEE/CVF International Conference on Computer Vision (ICCV)},
  year={2025}
}

@article{yang2025order,
  author = {Yang, Bo and Cao, Ying},
  title = {Order Matters: Learning Element Ordering for Graphic Design Generation},
  year = {2025},
  issue_date = {August 2025},
  publisher = {Association for Computing Machinery},
  address = {New York, NY, USA},
  volume = {44},
  number = {4},
  issn = {0730-0301},
  url = {https://doi.org/10.1145/3730858},
  doi = {10.1145/3730858},
  journal = {ACM Trans. Graph.},
  month = jul,
  articleno = {34},
  numpages = {16}
}

@inproceedings{peebles2023scalable,
  title={Scalable diffusion models with transformers},
  author={Peebles, William and Xie, Saining},
  booktitle={Proceedings of the IEEE/CVF international conference on computer vision},
  pages={4195--4205},
  year={2023}
}

@misc{brooks2023instructpix2pixlearningfollowimage,
      title={InstructPix2Pix: Learning to Follow Image Editing Instructions},
      author={Tim Brooks and Aleksander Holynski and Alexei A. Efros},
      year={2023},
      eprint={2211.09800},
      archivePrefix={arXiv},
      primaryClass={cs.CV},
      url={https://arxiv.org/abs/2211.09800},
}

@article{hertz2022prompt,
  title={Prompt-to-prompt image editing with cross attention control},
  author={Hertz, Amir and Mokady, Ron and Tenenbaum, Jay and Aberman, Kfir and Pritch, Yael and Cohen-Or, Daniel},
  journal={arXiv preprint arXiv:2208.01626},
  year={2022}
}

@inproceedings{bar2022text2live,
  title={Text2live: Text-driven layered image and video editing},
  author={Bar-Tal, Omer and Ofri-Amar, Dolev and Fridman, Rafail and Kasten, Yoni and Dekel, Tali},
  booktitle={European Conference on Computer Vision},
  pages={707--723},
  year={2022},
  organization={Springer}
}

@inproceedings{Zhang2023MagicBrush,
      title={MagicBrush: A Manually Annotated Dataset for Instruction-Guided Image Editing},
      author={Kai Zhang and Lingbo Mo and Wenhu Chen and Huan Sun and Yu Su},
      booktitle={Advances in Neural Information Processing Systems},
      year={2023}
}

@article{yu2024anyedit,
  title={AnyEdit: Mastering Unified High-Quality Image Editing for Any Idea},
  author={Yu, Qifan and Chow, Wei and Yue, Zhongqi and Pan, Kaihang and Wu, Yang and Wan, Xiaoyang and Li, Juncheng and Tang, Siliang and Zhang, Hanwang and Zhuang, Yueting},
  journal={arXiv preprint arXiv:2411.15738},
  year={2024}
}

@inproceedings{huang2024smartedit,
  title={Smartedit: Exploring complex instruction-based image editing with multimodal large language models},
  author={Huang, Yuzhou and Xie, Liangbin and Wang, Xintao and Yuan, Ziyang and Cun, Xiaodong and Ge, Yixiao and Zhou, Jiantao and Dong, Chao and Huang, Rui and Zhang, Ruimao and others},
  booktitle={Proceedings of the IEEE/CVF Conference on Computer Vision and Pattern Recognition},
  pages={8362--8371},
  year={2024}
}

@article{liu2025step1x-edit,
  title={Step1X-Edit: A Practical Framework for General Image Editing},
  author={Shiyu Liu and Yucheng Han and Peng Xing and Fukun Yin and Rui Wang and Wei Cheng and Jiaqi Liao and Yingming Wang and Honghao Fu and Chunrui Han and Guopeng Li and Yuang Peng and Quan Sun and Jingwei Wu and Yan Cai and Zheng Ge and Ranchen Ming and Lei Xia and Xianfang Zeng and Yibo Zhu and Binxing Jiao and Xiangyu Zhang and Gang Yu and Daxin Jiang},
  journal={arXiv preprint arXiv:2504.17761},
  year={2025}
}

@misc{wu2025qwenimagetechnicalreport,
      title={Qwen-Image Technical Report},
      author={Chenfei Wu and Jiahao Li and Jingren Zhou and Junyang Lin and Kaiyuan Gao and Kun Yan and Sheng-ming Yin and Shuai Bai and Xiao Xu and Yilei Chen and Yuxiang Chen and Zecheng Tang and Zekai Zhang and Zhengyi Wang and An Yang and Bowen Yu and Chen Cheng and Dayiheng Liu and Deqing Li and Hang Zhang and Hao Meng and Hu Wei and Jingyuan Ni and Kai Chen and Kuan Cao and Liang Peng and Lin Qu and Minggang Wu and Peng Wang and Shuting Yu and Tingkun Wen and Wensen Feng and Xiaoxiao Xu and Yi Wang and Yichang Zhang and Yongqiang Zhu and Yujia Wu and Yuxuan Cai and Zenan Liu},
      year={2025},
      eprint={2508.02324},
      archivePrefix={arXiv},
      primaryClass={cs.CV},
      url={https://arxiv.org/abs/2508.02324},
}

@misc{firered2026imageedit,
      title={FireRed-Image-Edit-1.1},
      author={{FireRed Team}},
      year={2026},
      howpublished={Hugging Face model},
      url={https://huggingface.co/FireRedTeam/FireRed-Image-Edit-1.1},
      note={Based on the FireRed-Image-Edit-1.0 Technical Report, arXiv:2602.13344},
}

@misc{schuhmann2022laion5bopenlargescaledataset,
      title={LAION-5B: An open large-scale dataset for training next generation image-text models},
      author={Christoph Schuhmann and Romain Beaumont and Richard Vencu and Cade Gordon and Ross Wightman and Mehdi Cherti and Theo Coombes and Aarush Katta and Clayton Mullis and Mitchell Wortsman and Patrick Schramowski and Srivatsa Kundurthy and Katherine Crowson and Ludwig Schmidt and Robert Kaczmarczyk and Jenia Jitsev},
      year={2022},
      eprint={2210.08402},
      archivePrefix={arXiv},
      primaryClass={cs.CV},
      url={https://arxiv.org/abs/2210.08402},
}

@misc{kingma2022autoencodingvariationalbayes,
      title={Auto-Encoding Variational Bayes},
      author={Diederik P Kingma and Max Welling},
      year={2022},
      eprint={1312.6114},
      archivePrefix={arXiv},
      primaryClass={stat.ML},
      url={https://arxiv.org/abs/1312.6114},
}

@misc{li2024layoutditexploringcontentgraphicbalance,
      title={LayoutDiT: Exploring Content-Graphic Balance in Layout Generation with Diffusion Transformer},
      author={Yu Li and Yifan Chen and Gongye Liu and Fei Yin and Qingyan Bai and Jie Wu and Hongfa Wang and Ruihang Chu and Yujiu Yang},
      year={2024},
      eprint={2407.15233},
      archivePrefix={arXiv},
      primaryClass={cs.CV},
      url={https://arxiv.org/abs/2407.15233},
}

@inproceedings{Hsu-IJCAI2025-ScanandPrint,
  title={Scan-and-Print: Patch-level Data Summarization and Augmentation for Content-aware Layout Generation in Poster Design},
  author={Hsu, HsiaoYuan and Peng, Yuxin},
  booktitle={Proceedings of the International Joint Conference on Artificial Intelligence},
  year={2025}
}

@inproceedings{horita2024retrievalaugmented,
    title={{Retrieval-Augmented Layout Transformer for Content-Aware Layout Generation}},
    author={Daichi Horita and Naoto Inoue and Kotaro Kikuchi and Kota Yamaguchi and Kiyoharu Aizawa},
    booktitle={CVPR},
    year={2024}
}

@misc{liu2025unilaydiffunifieddiffusiontransformer,
      title={UniLayDiff: A Unified Diffusion Transformer for Content-Aware Layout Generation},
      author={Zeyang Liu and Le Wang and Sanping Zhou and Yuxuan Wu and Xiaolong Sun and Gang Hua and Haoxiang Li},
      year={2025},
      eprint={2512.08897},
      archivePrefix={arXiv},
      primaryClass={cs.CV},
      url={https://arxiv.org/abs/2512.08897},
}

@misc{tian2025relayoutintegratingrelationreasoning,
      title={ReLayout: Integrating Relation Reasoning for Content-aware Layout Generation with Multi-modal Large Language Models},
      author={Jiaxu Tian and Xuehui Yu and Yaoxing Wang and Pan Wang and Guangqian Guo and Shan Gao},
      year={2025},
      eprint={2507.05568},
      archivePrefix={arXiv},
      primaryClass={cs.CV},
      url={https://arxiv.org/abs/2507.05568},
}

@misc{hsu2025posterostructuringlayouttrees,
      title={PosterO: Structuring Layout Trees to Enable Language Models in Generalized Content-Aware Layout Generation},
      author={HsiaoYuan Hsu and Yuxin Peng},
      year={2025},
      eprint={2505.07843},
      archivePrefix={arXiv},
      primaryClass={cs.GR},
      url={https://arxiv.org/abs/2505.07843},
}

@inproceedings{wang2025sega,
  title={SEGA: A Stepwise Evolution Paradigm for Content-Aware Layout Generation with Design Prior},
  author={Wang, Haoran and Zhao, Bo and Wang, Jinghui and Wang, Hanzhang and Yang, Huan and Ji, Wei and Liu, Hao and Xiao, Xinyan},
  booktitle={Proc. IEEE/CVF Int. Conf. Comput. Vis. (ICCV)},
  pages={19321--19330},
  year={2025}
}

@misc{yang2025qwen3technicalreport,
      title={Qwen3 Technical Report},
      author={An Yang and Anfeng Li and Baosong Yang and Beichen Zhang and Binyuan Hui and Bo Zheng and Bowen Yu and Chang Gao and Chengen Huang and Chenxu Lv and Chujie Zheng and Dayiheng Liu and Fan Zhou and Fei Huang and Feng Hu and Hao Ge and Haoran Wei and Huan Lin and Jialong Tang and Jian Yang and Jianhong Tu and Jianwei Zhang and Jianxin Yang and Jiaxi Yang and Jing Zhou and Jingren Zhou and Junyang Lin and Kai Dang and Keqin Bao and Kexin Yang and Le Yu and Lianghao Deng and Mei Li and Mingfeng Xue and Mingze Li and Pei Zhang and Peng Wang and Qin Zhu and Rui Men and Ruize Gao and Shixuan Liu and Shuang Luo and Tianhao Li and Tianyi Tang and Wenbiao Yin and Xingzhang Ren and Xinyu Wang and Xinyu Zhang and Xuancheng Ren and Yang Fan and Yang Su and Yichang Zhang and Yinger Zhang and Yu Wan and Yuqiong Liu and Zekun Wang and Zeyu Cui and Zhenru Zhang and Zhipeng Zhou and Zihan Qiu},
      year={2025},
      eprint={2505.09388},
      archivePrefix={arXiv},
      primaryClass={cs.CL},
      url={https://arxiv.org/abs/2505.09388},
}

@misc{gupta2021layouttransformerlayoutgenerationcompletion,
      title={LayoutTransformer: Layout Generation and Completion with Self-attention},
      author={Kamal Gupta and Justin Lazarow and Alessandro Achille and Larry Davis and Vijay Mahadevan and Abhinav Shrivastava},
      year={2021},
      eprint={2006.14615},
      archivePrefix={arXiv},
      primaryClass={cs.CV},
      url={https://arxiv.org/abs/2006.14615},
}

@inproceedings{Hsu-2023-posterlayout,
    title={PosterLayout: A New Benchmark and  Approach for Content-Aware Visual-Textual Presentation Layout},
    author={HsiaoYuan Hsu and Xiangteng He and Yuxin Peng and Hao Kong and Qing Zhang},
    booktitle={Proceedings of IEEE Conference on Computer Vision and Pattern Recognition (CVPR)},
    year={2023},
    pages={6018-6026}
}

@misc{yang2026interil,
      title={Learning Interaction between Image and Layout Priors for Joint Image-Layout Generation in Design Templates},
      author={Shirong Yang and Bo Yang and Ying Cao},
      year={2026},
      eprint={2609.11519},
      archivePrefix={arXiv},
      primaryClass={cs.CV},
      url={https://arxiv.org/abs/2609.11519},
}

@inproceedings{hou2026humanaware,
  title     = {Human-aware Design Generation: Adding 3D Humans into Graphic Designs},
  author    = {Hou, Zijin and Cao, Ying},
  booktitle = {Proceedings of the 34th ACM International Conference on Multimedia},
  year      = {2026},
  doi       = {10.1145/3767308.3836019}
}

\end{document}


\bstctlcite{BSTcontrol}

\maketitle

\section{Intention Generation Prompt}
\label{sec:intention_prompt}

To generate textual \textit{design intentions} for each sample in the CGL~\cite{zhou2022composition} and Crello~\cite{yamaguchi2021canvasvae} datasets, we employ \textbf{Qwen3-VL}, a state-of-the-art multimodal vision--language model capable of reasoning over complex visual compositions. Our prompt design is adapted from COLE~\cite{jia2023cole}, but specifically tailored for the task of design-intention inference. Unlike the original COLE setup---which focuses on descriptive captioning---our formulation emphasizes the \emph{purpose, style, and mood} of each design while deliberately excluding any textual information contained within the image. This adjustment ensures that the generated intentions faithfully capture the visual semantics and aesthetic intent of each design, even for text-heavy advertising posters.

\paragraph{Prompt setup.}
We employ two coordinated prompts: a \textit{system prompt} that defines the model's analytical role and behavioral constraints, and a \textit{user prompt} that directs it to produce a concise, purpose-oriented description. The complete prompt configuration used with Qwen3-VL is shown below.

\begin{tcolorbox}[
  colback=purple!5,
  colframe=purple!50!black,
  fonttitle=\bfseries,
  coltitle=white,
  title=Intention Generation Prompt for Qwen3-VL,
  colbacktitle=purple!70!black,
  boxrule=1pt,
  arc=4pt,
  outer arc=4pt,
  left=4pt,
  right=4pt,
  top=6pt,
  bottom=6pt,
  toptitle=2pt,
  bottomtitle=2pt,
  breakable
]
\textbf{System Prompt:}

You are an excellent image analyst and capable of inferring a designer's intention from a visual design. You will be given a single commercial poster. Your task is to infer the underlying design intention behind the poster, not merely describe what is visible.

\textbf{Guidelines:}
\begin{itemize}[leftmargin=*]
  \item Focus on design purpose, style, and mood (e.g., ``promote a summer sale'', ``celebrate Mother's Day'').
  \item Do \textbf{not} mention any text visible in the image.
  \item Do \textbf{not} list specific text contents, URLs, or prices.
  \item Keep the intention short (no more than 20 words).
  \item If necessary, make a reasonable guess based on the visual theme.
\end{itemize}

\textbf{User Prompt:}

Describe the likely *design intention* behind this image, focusing on its visual purpose and aesthetic style. Do \textbf{not} repeat any text visible in the image. Keep it short and clear.
\end{tcolorbox}

The generated intention summarizes the purpose, style, and mood of each design
while excluding visible textual content, and is used as the textual condition
for training the CELG model.

\section{Implementation Details}
\label{sec:implementation_details}

\noindent\textbf{Layout Representation.}
Each layout is represented as a sequence of up to 10 elements, where each element is defined by five attributes: category, $x_1$, $y_1$, $x_2$, $y_2$ (normalized bounding box coordinates). Elements are padded to a fixed sequence length, covering the vast majority of samples in the CGL dataset.

\noindent\textbf{Layout VAE (Stage 1).}
We use a Transformer-based $\beta$-VAE ($\beta = 5 \times 10^{-4}$) to encode layout tokens into an 8-dimensional latent space per element. The encoder consists of 10 Transformer layers (hidden size 512, 8 attention heads) that produce per-element mean and log-variance vectors. The decoder mirrors this architecture and reconstructs both element categories (via cross-entropy loss) and bounding box coordinates (via L1/L2 regression loss). Training uses AdamW (lr=$10^{-4}$).

\noindent\textbf{Layout DiT (Stage 2).}
A DiT-based transformer serves as the layout flow matching backbone, consisting of 28 transformer blocks with hidden size 512 and 8 attention heads. We train with flow matching ($T{=}1000$ timesteps) using velocity prediction $v = \epsilon - l_0$. The layout latents are normalized by the standard deviation of the first training batch before feeding to the DiT. The timestep provides global conditioning through adaptive layer normalization, while the foreground requirement is injected per element as described below. Training uses AdamW (lr=$10^{-4}$).

\noindent\textbf{Slot-Aligned Foreground Requirement Conditioning.}
The layout VAE preserves the element axis: its noisy latent has shape
$l_\sigma\in\mathbb{R}^{B\times S\times 8}$, with $S=10$, and row $j$
represents the element slot being denoised at that position. After the input
projection, the DiT token sequence has shape
$x\in\mathbb{R}^{B\times S\times 512}$. We therefore inject the requirement
for element $j$ directly into row $j$. The requirement triples are serialized
into the corresponding element slots; unused padded slots take the null
entries and remain masked.

We use three independent embedding tables for category, footprint, and
aspect. The footprint and aspect tables contain five and four class entries,
respectively, and every table has one additional \emph{null} entry. For a
nonempty requirement, the slot condition and initial DiT token are
\begin{align}
    e_j &=
    E_{\mathrm{cat}}(c_j)
    + E_{\mathrm{foot}}(f_j)
    + E_{\mathrm{asp}}(a_j),\\
    x_j^{(0)} &=
    \operatorname{FC}_{\mathrm{in}}(l_{\sigma,j})
    + p_j + e_j,
\end{align}
where $p_j$ is the positional embedding. The requirement is added once,
before the first DiT block. Unlike the timestep embedding, which is a single
$B\times512$ vector that modulates all tokens through adaptive layer
normalization, $(c_j,f_j,a_j)$ varies across the $S$ element slots and must
remain on the token axis.

The conditioned tokens then pass through the 28 self-attention blocks.
Self-attention allows slots with different requirements to coordinate their
sizes and positions, while the final layer predicts one denoised latent per
slot. The frozen layout VAE decodes these latents into element categories and
boxes. During training, $(c_j,f_j,a_j)$ is derived from the corresponding
ground-truth element, and the three embedding tables are learned through the
same flow-matching velocity loss as the layout denoiser. Thus, their vectors
act as learned, composable directions that steer each slot toward the
requested semantic and geometric region of the layout latent space.

For classifier-free training, we replace the complete requirement with null
entries with probability $0.1$. Unconstrained CELG
($\mathcal{C}=\varnothing$) uses this all-null branch directly. At inference,
the strength of a nonempty requirement can be adjusted with velocity
guidance,
\begin{equation}
    \hat{v}
    = \hat{v}_{\varnothing}
    + s\left(\hat{v}_{\mathcal{C}}-\hat{v}_{\varnothing}\right),
\end{equation}
where $s$ trades diversity against requirement satisfaction.

\noindent\textbf{Image Editing Backbone.}
The image editing branch uses FireRed-Image-Edit-1.1~\cite{firered2026imageedit}, a pretrained 20B flow-matching editor built on the Qwen-image architecture, with 60 dual-stream blocks (hidden dimension 3072, 24 attention heads). Given a coarse background image, we encode it using a VAE with 8$\times$ spatial compression and 16 latent channels. The latents are spatially packed with patch size 2, yielding 1024 tokens per image. The transformer input consists of 2048 tokens: 1024 noisy latent tokens concatenated with 1024 reference (source image) tokens. The edit instruction is encoded by the editor's pretrained text encoder and supplied as text conditioning.

\noindent\textbf{Cross-Modal Communication (Stage 3).}
As described in Section 3 of the main paper, we couple the image and layout branches through a warm-up phase followed by three rounds of interleaved bidirectional exchange.

\noindent\textit{Phase 1 (Warm-up).}
The image transformer runs its first 20 blocks (0--19) and the layout DiT runs its first 15 blocks (0--14) independently, building modality-specific representations before any information exchange.

\noindent\textit{Rounds 1--3 (Bidirectional Exchange).}
Each round first passes image context to the layout branch and then returns the updated layout context to the image branch:

\begin{itemize}[leftmargin=*]
  \item \textbf{Image-to-layout bridge:} FireRed image features ($\mathbb{R}^{B \times 1024 \times 3072}$) are projected to the 512-dimensional layout space and supplied as keys and values in cross-attention, with the layout tokens serving as queries.
  \item \textbf{Layout-to-image bridge:} After the corresponding layout segment is processed, its features ($\mathbb{R}^{B \times 10 \times 512}$) are projected to the 3072-dimensional image space and supplied as keys and values, with the image tokens serving as queries.
\end{itemize}

In Round 1, image features after blocks 0--19 condition layout blocks 15--19, whose output is returned before image blocks 20--33. Round 2 uses the image representation after blocks 20--33 to condition layout blocks 20--23, then feeds the updated layout representation back before image blocks 34--45. Round 3 repeats the process between layout blocks 24--27 and image blocks 46--59, using the image representation produced by blocks 34--45.

The three rounds collectively consume all 60 image transformer blocks and all 28 layout DiT blocks, with no leftover tail. All communication blocks use zero initialization for output projections, ensuring training starts from pretrained model behavior.

\noindent\textit{Trainable Parameters.}
The FireRed transformer and VAE are fully frozen throughout Stage 3. Layout DiT blocks 0--23 are also frozen. Only the following parameters are updated:
\begin{itemize}[leftmargin=*]
  \item three image-to-layout cross-attention bridges;
  \item three layout-to-image cross-attention bridges;
  \item Layout DiT blocks 24--27 and the final projection layer.
\end{itemize}
Training uses AdamW (lr=$1\times10^{-4}$, gradient accumulation=4).

\noindent\textbf{Inference Configuration.}
We use the Euler sampler (50 steps) for both layout and image generation with shared timestep scheduling. Classifier-free guidance (scale 4.0) is applied to the image branch with norm rescaling. Both branches operate synchronously, exchanging features at the three interleaved communication points described above.

\section{Requirement-Conditioned Baseline Adaptations}
\label{sec:requirement_conditioned_baselines}

For a controlled comparison under the CELG formulation, we adapt three
content-aware layout generators to accept the same foreground requirement
$\mathcal{C}=\{(c_i,f_i,a_i)\}_{i=1}^{M}$ used by our method. Here $c_i$
denotes the semantic category, $f_i$ the footprint class, and $a_i$ the
aspect class of element $i$. The adaptations preserve the original visual
conditioning and layout-generation mechanisms of each baseline. These
adaptations are activated for Requirement-Conditioned CELG
($\mathcal{C}\neq\varnothing$); when $\mathcal{C}=\varnothing$, the original
unconstrained baseline is used.

\paragraph{Common $\mathcal{C}$-aware edit instruction.}
For the Requirement-Conditioned comparison reported in Panel B of the
main-paper comparison table, Qwen3-VL receives $I_{\mathrm{src}}$, the design intention $u$,
and a textual description of the requested element categories, footprints,
and aspect classes in $\mathcal{C}$, and produces one edit
instruction $\mathcal{T}_{\mathcal{C}}$. This exact instruction is supplied to
the common FireRed-Image-Edit-1.1 backbone for all three sequential baselines
and to the image branch of our joint model. The sequential baselines therefore
share the same edited canvas and subsequently condition their respective
layout generators on $\mathcal{C}$. No method receives a target layout or
ground-truth box coordinates. This protocol equalizes the information and
editing backbone available to all methods; the remaining distinction is
whether layout-to-image feedback is available during generation.

The exact instructions used for the three cases in the main-paper qualitative
comparison are provided below.

\begin{tcolorbox}[
  colback=cyan!4,
  colframe=cyan!45!black,
  fonttitle=\bfseries,
  title=Case 1: Travel Organizer,
  boxrule=0.8pt, arc=3pt,
  left=4pt, right=4pt, top=4pt, bottom=4pt,
  breakable
]
Transform this product collage into a clean travel-themed promotional scene.
Keep the turquoise organizer unchanged and clearly visible. Place it on a
table near an open suitcase, map, sunglasses, and camera, with a bright
tropical beach view in the background. Remove size labels and extra product
photos, and create a vibrant, practical vacation atmosphere.
\end{tcolorbox}

\begin{tcolorbox}[
  colback=orange!5,
  colframe=orange!55!black,
  fonttitle=\bfseries,
  title=Case 2: Whisky,
  boxrule=0.8pt, arc=3pt,
  left=4pt, right=4pt, top=4pt, bottom=4pt,
  breakable
]
Turn this collage of whisky photos into a single premium advertising scene.
Keep the whisky glasses and bottle elements recognizable, and combine them
into one dark, elegant composition with warm lighting and a cinematic mood.
Remove the collage layout and white borders, add a refined bar-like setting,
and create a luxurious atmosphere with clear space around the products.
\end{tcolorbox}

\begin{tcolorbox}[
  colback=green!4,
  colframe=green!45!black,
  fonttitle=\bfseries,
  title=Case 3: Office Chair,
  boxrule=0.8pt, arc=3pt,
  left=4pt, right=4pt, top=4pt, bottom=4pt,
  breakable
]
Transform this product collage into a clean lifestyle advertisement scene.
Keep the pink ergonomic chair unchanged and clearly recognizable. Remove the
extra inset views and place the chair in a bright, cozy room with soft natural
light, a desk, shelves, and simple decor. Create a friendly, modern atmosphere
while keeping the focus on the chair.
\end{tcolorbox}

\paragraph{Why RALF is not adapted.}
RALF~\cite{horita2024retrievalaugmented} is evaluated only under
Unconstrained CELG. Its constraints are supplied as a discrete token sequence
through a dedicated user-constraint encoder, so introducing footprint and
aspect classes requires extending the constraint vocabulary. New entries
shift the existing token indices and therefore invalidate the released
checkpoint. In addition, RALF enforces its size-conditioned task by copying
the given size tokens directly into the decoded output, which is possible
only because the condition and the prediction occupy the same vocabulary
positions. A $(f_i,a_i)$ pair instead specifies a \emph{set} of admissible
width--height bins rather than a single one, so satisfying it would require a
new joint decoding restriction over the two coordinates. Adapting RALF is
thus a retraining rather than a fine-tuning exercise, and a RALF retrained
from scratch would not be comparable on equal terms with the other three
baselines, which are adapted from their released weights. We therefore keep
RALF in the unconstrained comparison only, where it is evaluated exactly as
published.

\subsection{Requirement-Conditioned Scan-and-Print}
\label{sec:scanprint_req}

To isolate the contribution of background adaptation, we equip
Scan-and-Print~\cite{Hsu-IJCAI2025-ScanandPrint} with $\mathcal{C}$ while
keeping its canvas fixed. Scan-and-Print encodes a poster canvas with a ViT,
retains only the $K$ most salient patches, and autoregressively decodes a
layout token sequence that cross-attends to these patches. We inject
$\mathcal{C}$ on the encoder side: each triple $(c_i,f_i,a_i)$ is mapped to
a condition token
\begin{equation}
    r_i = E_{c}(c_i) + E_{f}(f_i) + E_{a}(a_i),
\end{equation}
where $E_{c}$, $E_{f}$, and $E_{a}$ are learned embedding tables. The
resulting set $\{r_i\}_{i=1}^{M}$ is concatenated with the retained patch
tokens to form the decoder memory.

We deliberately treat $\mathcal{C}$ as an \emph{unordered set} rather than
aligning $r_i$ with sequence positions. Scan-and-Print serializes each
element into two spatially reordered endpoints, so the correspondence
between sequence positions and elements is determined only during decoding;
cross-attention learns this correspondence implicitly. Because $f_i$ and
$a_i$ are derived from ground-truth boxes, the variant requires no
additional annotation and otherwise retains the original training recipe.
We can additionally mask category logits according to the remaining
requirement budget during decoding, thereby enforcing the category multiset
and element count $M$ without retraining. This baseline satisfies
$\mathcal{C}$ only when the given canvas admits a feasible layout, isolating
the limitation that CELG is designed to address.

\subsection{Requirement-Conditioned PosterLlama}
\label{sec:posterllama_req}

PosterLlama~\cite{seol2024posterllama} serializes a layout into HTML and
autoregressively decodes it with a code language model, so a requirement can
be supplied without an architectural change. We extend its element template
with two attributes carrying the footprint and aspect classes:
\begin{center}
\small
\texttt{<rect data-category="\{$c_i$\}", data-footprint="\{$f_i$\}",}\\
\texttt{data-aspect="\{$a_i$\}", x="<M>", y="<M>", width="<M>", height="<M>"/>}
\end{center}
The coordinate fields remain masked and are predicted by the model. The
target sequence retains the original output template, leaving decoding and
evaluation unchanged.

For each element, we compute the normalized area
$A_i=(x_2-x_1)(y_2-y_1)$ and aspect ratio
$R_i=(x_2-x_1)/(y_2-y_1)$ from its ground-truth box, and assign $f_i$ and
$a_i$ by quantile binning over the training set. This produces five
footprint classes and four aspect classes with approximately balanced
populations. The resulting conditional task,
$\mathcal{C}\!\rightarrow\!\mathcal{L}$, generalizes PosterLlama's existing
category- and size-conditioned tasks by replacing exact dimensions with
coarse classes.

Because $(f_i,a_i)$ are derived from the ground truth, training exclusively
on perfectly consistent pairs could encourage the model to invert the bins
and reproduce the reference box instead of treating the requirement as a
specification. We therefore perturb each requirement to an adjacent bin with
probability $p$ during training. The new task is mixed with PosterLlama's
original conditional tasks, and we fine-tune the released second-stage
checkpoint using LoRA with rank $r=64$ for a small number of epochs. The
vision encoder and projection layer remain frozen.

\subsection{Requirement-Conditioned LayoutDiT}
\label{sec:layoutdit_req}

For LayoutDiT~\cite{li2024layoutditexploringcontentgraphicbalance}, we inject
$\mathcal{C}$ into the layout denoiser as a per-element conditioning signal.
Because each requirement triple corresponds to one element slot, we align
the requirement with the layout sequence by position instead of attending
over it as an unordered set.

Each triple is mapped to a vector by three learned embedding tables,
\begin{equation}
    \mathbf{e}_i =
    E_{\text{cat}}(c_i) + E_{\text{foot}}(f_i) + E_{\text{asp}}(a_i),
\end{equation}
which is added to the corresponding layout token before the first
Transformer block:
\begin{equation}
    \mathbf{h}_i^{(0)}
    = \operatorname{MLP}(\mathbf{l}_i^{t}) + \mathbf{p}_i + \mathbf{e}_i,
\end{equation}
where $\mathbf{l}_i^{t}$ is the noisy layout token at step $t$ and
$\mathbf{p}_i$ is its positional embedding. We re-inject
$\mathbf{e}_i$ at the decoder input so that the requirement remains visible
to deeper layers.

The semantic category $c_i$ is additionally enforced as a hard constraint.
Following masked inpainting for conditional layout generation, its one-hot
channels remain noise-free during the forward process and are reset to the
given values after every reverse step. Consequently,
$E_{\text{cat}}$ supplies context for the footprint and aspect conditions,
while the network learns to satisfy the two coarse geometric constraints.

Each embedding table includes a \emph{null} entry. During training, we
replace the entire requirement with null triples with probability $p=0.1$,
allowing one model to support both unconstrained and
requirement-conditioned generation and enabling classifier-free guidance:
\begin{equation}
    \hat{\boldsymbol{\epsilon}}
    = \hat{\boldsymbol{\epsilon}}_{\varnothing}
    + s\left(
        \hat{\boldsymbol{\epsilon}}_{\mathcal{C}}
        - \hat{\boldsymbol{\epsilon}}_{\varnothing}
      \right),
\end{equation}
where $s$ trades layout diversity against requirement satisfaction. As
above, the bin edges for $f_i$ and $a_i$ are empirical quantiles of the
training split, producing five footprint and four aspect classes with
approximately balanced populations.

\section{Unconstrained CELG Qualitative Results}
\label{sec:unconstrained_qualitative}

\begin{figure*}[t]
  \centering
  {\scriptsize
  \setlength{\tabcolsep}{0pt}
  \begin{tabular}{@{}
    p{0.115\textwidth}p{0.128\textwidth}p{0.136\textwidth}%
    p{0.126\textwidth}p{0.123\textwidth}p{0.123\textwidth}%
    p{0.128\textwidth}p{0.121\textwidth}@{}}
  \centering Source Asset &
  \centering Design Intention &
  \centering Edit Instruction &
  \centering PosterLlama &
  \centering Scan-and-Print &
  \centering RALF &
  \centering LayoutDiT &
  \centering Ours
  \end{tabular}}
  \vspace{2pt}
  \includegraphics[width=\textwidth,trim=0 0 0 20,clip]{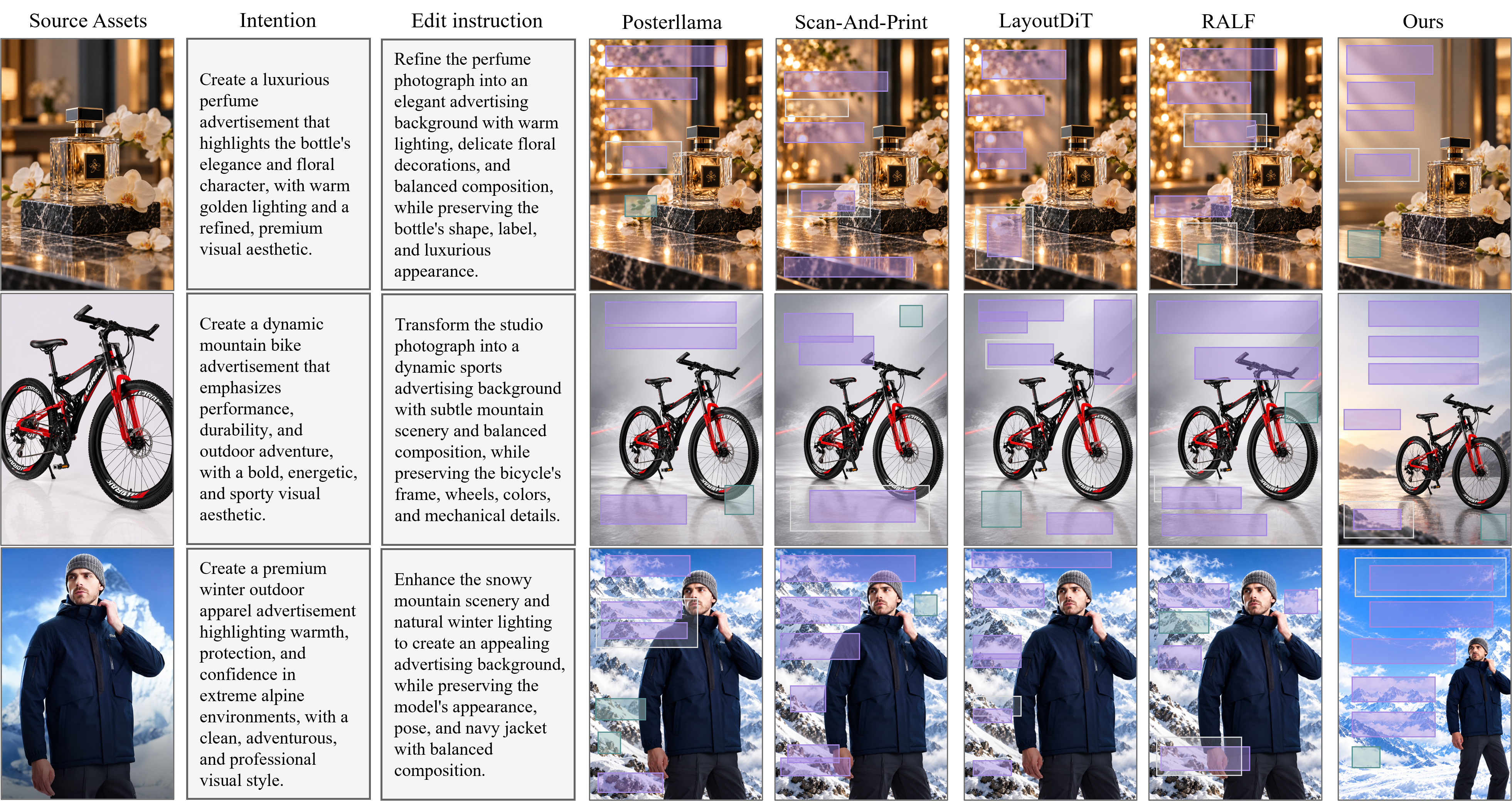}
  \caption{Unconstrained CELG qualitative comparison corresponding to Panel A
  of Table~III in the main paper. Each row shows a source asset, design
  intention, edit instruction, and the results of four sequential CLG
  baselines and Ours. The sequential baselines share the same edited canvas,
  whereas Ours jointly adapts the canvas and layout. Purple boxes denote text,
  white boxes denote underlays, and cyan boxes denote logos.}
  \label{fig:unconstrained_qualitative}
\end{figure*}

Figure~\ref{fig:unconstrained_qualitative} complements the quantitative
results for Unconstrained CELG, where no foreground requirement is provided
and each layout generator determines the foreground elements itself. The
sequential baselines place their layouts on a canvas fixed before layout
generation, which can leave foreground elements competing with the perfume
bottle, bicycle, or model. Our method instead adjusts the visual composition
during joint generation: it places the principal subject toward one side and
creates larger, more continuous regions for foreground content. These examples
support the quantitative finding that joint canvas--layout generation provides
better overall quality than sequentially editing the image and then generating
the layout, even without an explicit foreground requirement.

\section{SCE Prompts for Instruction Generation}
\label{sec:sce_prompts}

As described in Section~3.2 of the main paper, the SCE pipeline constructs training triplets $(I_\text{src}, I_\text{tgt}, \mathcal{T})$ guided by three design observations: \textit{source diversity}, \textit{limited editing degree}, and \textit{rich editing dimensions} (spanning color, texture, shape, and object-level manipulation). Two \textbf{Qwen3-VL} prompts are used to generate the two types of instructions involved.

\paragraph{Recovery Instruction ($\mathcal{T}_r$).}
In the \textit{Compare} stage, the VLM receives the target design $I_\text{tgt}$ and its retrieved reference $I_\text{ref}$, and generates a recovery instruction $\mathcal{T}_r$ describing how to transform $I_\text{tgt}$ toward $I_\text{ref}$. The prompt directs the model to identify the single most significant visual change---either a shape-level edit (geometry, object identity, semantic content) or a color-level edit (tone, lighting, color identity)---consistent with the rich editing dimensions in the main paper.

\begin{tcolorbox}[
  colback=blue!5,
  colframe=blue!40!black,
  fonttitle=\bfseries,
  coltitle=white,
  title=Recovery Instruction Prompt ($\mathcal{T}_r$),
  colbacktitle=blue!60!black,
  boxrule=1pt, arc=4pt, outer arc=4pt,
  left=4pt, right=4pt, top=6pt, bottom=6pt,
  toptitle=2pt, bottomtitle=2pt,
  breakable
]
\textbf{System Prompt:}

You are an expert assistant specializing in design-level image editing instructions. Your task is to describe transformations between two images. Visual edits fall into two main categories:
\begin{itemize}[leftmargin=*]
  \item \textbf{Shape Variation:} Changes in form, geometry, or semantic content of elements. Examples: posture/pose changes (sleeping cat $\rightarrow$ sitting cat), element augmentation (adding wings to a car), object replacement/distortion (circle $\rightarrow$ star, box $\rightarrow$ bottle).
  \item \textbf{Color Variation:} Changes in aesthetic tone, lighting, or color identity. Examples: background recoloring (red $\rightarrow$ blue), tonal/lighting shift (daylight $\rightarrow$ night), local element recoloring (shirt yellow $\rightarrow$ green).
\end{itemize}
You must always describe the most significant SINGLE variation between the images. Keep instructions precise, professional, and unambiguous.

\textbf{User Prompt:}

Analyze the two images. Identify the most significant single visual change (Shape Variation or Color Variation). Write one direct recovery instruction describing how to transform the first image into the second. Do not mention multiple changes. Use concise, professional language suitable for design editing.
\end{tcolorbox}

\paragraph{Edit Instruction ($\mathcal{T}$).}
After the \textit{Edit} stage applies $\mathcal{T}_r$ to produce $I_\text{src}$, a second VLM call compares $I_\text{src}$ (BEFORE) and $I_\text{tgt}$ (AFTER) to generate the forward edit instruction $\mathcal{T}$ describing the transformation from source to target. This instruction forms the textual supervision signal used to train and condition the image editing branch.

\begin{tcolorbox}[
  colback=teal!5,
  colframe=teal!40!black,
  fonttitle=\bfseries,
  coltitle=white,
  title=Edit Instruction Prompt ($\mathcal{T}$),
  colbacktitle=teal!60!black,
  boxrule=1pt, arc=4pt, outer arc=4pt,
  left=4pt, right=4pt, top=6pt, bottom=6pt,
  toptitle=2pt, bottomtitle=2pt,
  breakable
]
\textbf{System Prompt:}

You are an expert assistant specializing in design-level image editing instructions. You will always receive two images in a fixed order:
\begin{itemize}[leftmargin=*]
  \item The FIRST image represents the original (BEFORE edit).
  \item The SECOND image represents the result (AFTER edit).
\end{itemize}
Your task is to describe the transformation needed to convert the FIRST image into the SECOND. Visual edits fall into two main categories:
\begin{itemize}[leftmargin=*]
  \item \textbf{Shape Variation:} Changes in form, geometry, or semantic content. Examples: posture/pose changes (sleeping cat $\rightarrow$ sitting cat), adding/removing elements, object replacement.
  \item \textbf{Color Variation:} Changes in tone, lighting, or color identity. Examples: recoloring (red $\rightarrow$ blue), tone shift (daylight $\rightarrow$ night).
\end{itemize}
Always output a single concise instruction describing this transformation. Never describe the second image directly---only the edit operation.

\textbf{User Prompt:}

Analyze the two provided images:
\begin{itemize}[leftmargin=*]
  \item The FIRST image shows the original design (before edit).
  \item The SECOND image shows the modified design (after edit).
\end{itemize}
Write ONE precise, professional instruction describing how to transform the FIRST image into the SECOND. Use clear verbs like `change', `replace', `add', `remove', `recolor', `adjust'. Do not mention both images explicitly---describe only the editing operation.
\end{tcolorbox}

\section{CGL-Held-Out Evaluator Training}
\label{sec:liae_dataset}

We adopt TemplateFID and TemplateCLIP from InterIL~\cite{yang2026interil}.
Because CGL is the target benchmark in this work~\cite{zhou2022composition},
we retrain the evaluator checkpoints without using any CGL samples. The
evaluator architectures and training objectives otherwise follow InterIL.

To construct a sufficiently large CGL-held-out training corpus, we combine
templates from PKU~\cite{Hsu-2023-posterlayout},
WebDesign~\cite{weng2024desigen},
GenPoster-100K~\cite{wang2025sega}, and
Crello~\cite{yamaguchi2021canvasvae}. Their annotations are remapped to four
shared categories: logo/image, text, underlay/background, and
embellishment/CTA. We also discard templates containing more than 10 layout
elements, matching the maximum layout length used in our experiments. The
resulting corpus contains 148{,}529 training and 19{,}397 validation
templates, with 84{,}936, 44{,}661, 11{,}198, and 7{,}734 training templates
contributed by GenPoster-100K, WebDesign, Crello, and PKU, respectively.

Following InterIL, we train the layout VAE, TemplateAE, and TemplateCLIP on
this corpus. The layout encoder defines the external feature space used for
L-FID, TemplateAE provides the joint background--layout representation used
for TemplateFID, and TemplateCLIP evaluates the alignment between the
generated template and its design intention. All evaluator checkpoints are
trained independently and share no weights with CELG or the baseline models.

\section{Human Evaluation of CELG Outputs}
\label{sec:pairwise_user_study}

We conduct an anonymous human evaluation comparing Ours
($+\mathcal{C}$) with three requirement-conditioned CLG baselines:
LayoutDiT ($+\mathcal{C}$), Scan-and-Print ($+\mathcal{C}$), and
PosterLlama ($+\mathcal{C}$). These are the same adapted baselines evaluated
in Panel B of the main-paper comparison table: every method receives the same source
asset, design intention, and nonempty foreground requirement
$\mathcal{C}$.

\noindent\textbf{Participants and assignment.}
We recruit 34 participants. Each completes 21 evaluated trials and one
quality-control trial that is excluded from the reported results. The
evaluated trials are evenly divided among the three comparators, so every
participant assesses seven LayoutDiT, seven Scan-and-Print, and seven
PosterLlama results. This produces 238 comparisons per baseline. We use 30
test cases, with each baseline--case pair receiving seven or eight independent
judgments. A participant never sees the same source case more than once. All
selected cases require preservation of the primary product's identity and
non-targeted appearance.

\noindent\textbf{Two-stage interface.}
For each trial, the page header shows the source asset, the design intention,
and a natural-language description of the requested elements in $\mathcal{C}$
rather than its tuple-based representation. For example:
\begin{quote}
\small
\textbf{Required foreground:}\\
1 hero landscape text;\\
1 medium landscape text;\\
1 small square logo.
\end{quote}
The two candidate methods are anonymized as A and B. Their left--right
positions are randomized for every trial and then held fixed across its two
stages; method identities are never shown.

\noindent\textit{Canvas-level stage.}
Participants view the source asset, design intention, foreground requirement,
and the two anonymized edited canvases $\tilde I_A$ and $\tilde I_B$, but not
the generated layouts or resulting templates. They independently rate both
canvases on the three five-point scales described below. After submitting
these ratings, participants proceed to the next stage; canvas-level responses
cannot be revised after the resulting templates are revealed.

Because the three CLG baselines use the same
$\mathcal{C}$-aware instruction and image editor, their edited canvas is
identical for a given source. We therefore pool it as the
\textit{Sequential Editing} result rather than treating its repetitions as three
different methods. The layout-blind first stage ensures that these pooled
canvas scores cannot be influenced by the quality of the baseline layout with
which the canvas is subsequently paired.
\begin{itemize}[leftmargin=*]
    \item \textbf{Image Editing Quality} rates visual realism, absence of
    artifacts, and consistency with the design intention, from 1
    (\textit{bad}) to 5 (\textit{excellent}).
    \item \textbf{Product Preservation} rates how faithfully the result
    preserves the identity and non-targeted visual details of the source
    product, from 1 (\textit{not preserved}) to 5 (\textit{fully preserved}).
    \item \textbf{Canvas Adaptation Utility} rates whether the edit makes the
    canvas more suitable for the intended design and foreground requirement,
    from 1 (\textit{harmful}) to 5 (\textit{substantially beneficial}).
\end{itemize}
For each criterion, we report the mean five-point rating for Ours and
\textit{Sequential Editing}.

\noindent\textit{Template-level stage.}
After the canvas ratings are locked, the two resulting templates
$(\tilde I_A,\mathcal{L}_A)$ and
$(\tilde I_B,\mathcal{L}_B)$ are revealed. Participants compare them under the
following criteria:
\begin{itemize}[leftmargin=*]
    \item \textbf{Layout Quality} considers balance, alignment, spacing,
    readability, element collision, and accommodation of the foreground
    requirement.
    \item \textbf{Holistic Quality} considers whether the edited canvas and
    foreground layout form a coherent and professional template.
\end{itemize}
Each criterion has three response options: \textit{A better}, \textit{Tie},
and \textit{B better}. After undoing the randomized A/B assignment, we report
Win/Tie/Loss percentages for Ours against each comparator.

\bibliographystyle{IEEEtran}
\bibliography{main}